\documentclass[11pt]{amsart}
\usepackage{graphicx,xcolor,verbatim}
\usepackage{amssymb,amsfonts,amsmath}
\usepackage{graphics,epsfig}
\usepackage{graphicx}
\usepackage{latexsym, arcs}

\DeclareSymbolFont{yhlargesymbols}{OMX}{yhex}{m}{n}
\DeclareMathAccent{\wideparen}{\mathord}{yhlargesymbols}{"F3}

\newcommand{\be}{\begin{equation}}
\newcommand{\ee}{\end{equation}}

\begin{document}

\title[Iron rings formation]{A Mathematical and Experimental Study on Iron Rings Formation in Porous Stones}

\author[R. Reale, L. Campanella, et alia]{Rita Reale, Luigi Campanella, \\ Maria Pia Sammartino, Giovanni Visco \\ {\rm Chemistry  Department} \\ {\rm Sapienza University of Rome} \\ \\ Gabriella Bretti, Maurizio Ceseri, \\ Roberto Natalini, Filippo Notarnicola \\ {\rm Istituto per le Applicazioni del Calcolo ``M. Picone''} \\ {\rm Consiglio Nazionale delle Ricerche}}

\begin{abstract}

In this interdisciplinary paper, we study the formation of iron precipitates - the so-called Liesegang rings - in Lecce stones in contact with iron source. These phenomena are responsible of exterior damages of lapideous artifacts, but also in the weakening of their structure. They originate in presence of water, determining the flow of carbonate compunds mixing with the iron ions and then, after a sequence of reactions and precipitation, leading to the formation of Liesegang rings. In order to model these phenomena observed in situ and in laboratory experiments, we propose a modification of the classical Keller-Rubinow model and show the results obtained with some numerical simulations, in comparison with the experimental tests. Our model is of interest for a better understanding of damage processes in monumental stones. 
\end{abstract}
\maketitle

\section{Introduction}
\begin{figure}
\centering
\includegraphics[height=7 cm, width=4cm]{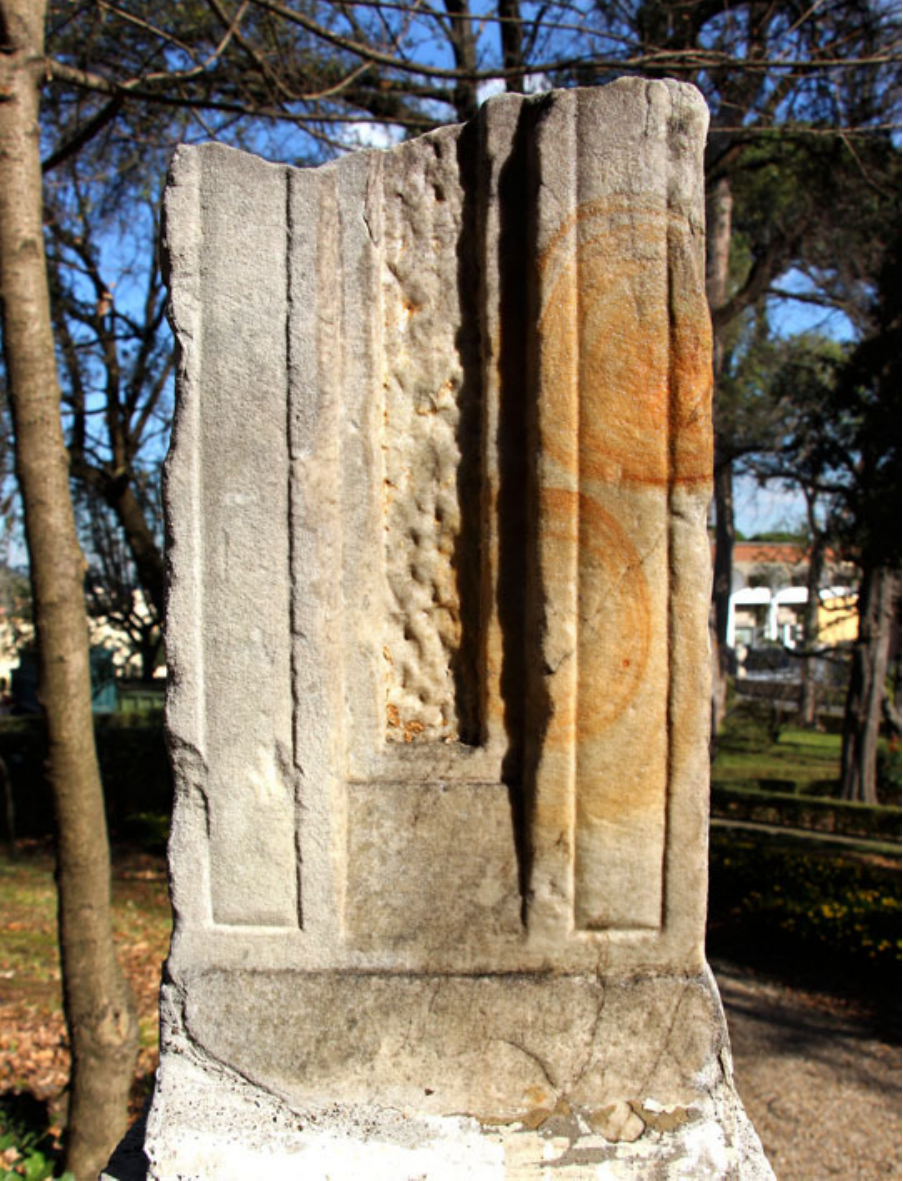}\quad
\includegraphics[height=7 cm, width=4cm]{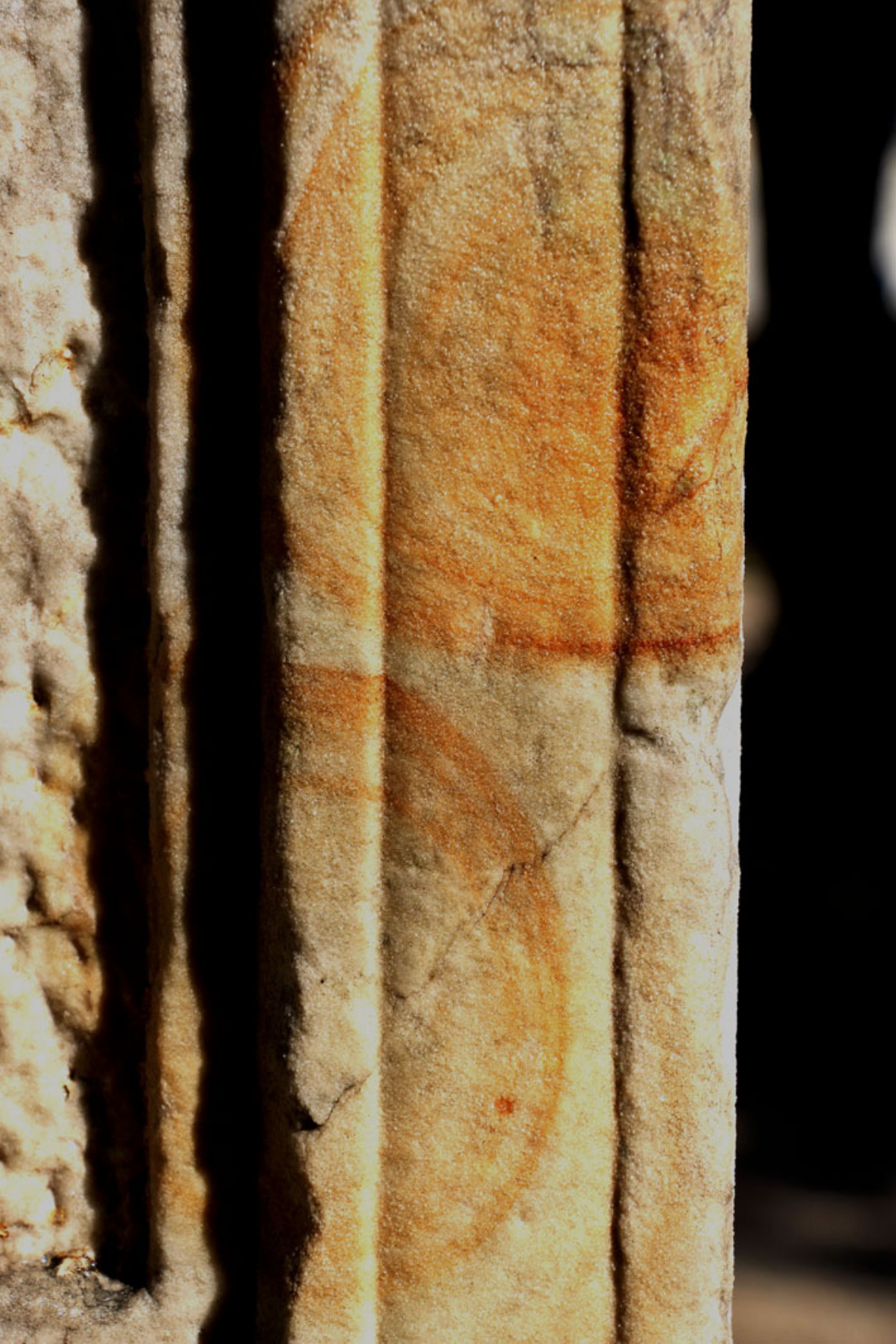}
\caption{Stones presenting iron rings. Vatican Gardens.}  \label{fig:colonna}   
\end{figure}

Among the numerous conservative issues regarding stone artifacts, the presence of stains caused by metallic compounds, in particular iron ones, results to be one of the deeply felt problem, both from the aesthetic aspect and the chemical-physical damage of substrate \cite{b, w}.
Iron, in fact, tends to oxidize rapidly, giving rise to colloidal hydrate oxide, that, in contact with porous materials, such as carbonate stones, may diffuse from their source (as an example, iron linchpin) to the near stone mass, often with particular shapes, such as Liesegang rings \cite{Liesegang} (see Figure \ref{fig:colonna}).
The formation of Liesegang rings represent a phenomenon of periodic precipitation, observed both in simulated chemical systems constituted by ions scattered in a colloidal medium in gel state, and in natural systems (sedimentary rocks). Such phenomenon causes the formation of coloured ring bands more or less regular.\\

\paragraph{\bf Goal of the paper.}
The study presented in this paper is devoted to the understanding of the phenomenon of iron Liesegang rings formation in stones and other materials and its mathematical modeling, in order to be able to improve our understanding of these phenomena and make predictions about the chemical damage of porous materials. To this aim, we performed an interdisciplinary study, both experimental and numerical. First, laboratory experiments were conducted on host porous materials. The first experiments were conducted on Lecce stone as host material, while a second experiment was performed on petri dish filled with colloidal solution of agar gel and sodium hydroxide. Both materials were put in contact with a iron source, in order to produce  the rings. Looking at the results obtained in laboratory,  numerical simulations were performed based on a modification of the  Keller-Rubinow model \cite{KR}, by using suitable parameters. The results obtained numerically show that the model is able to reproduce the main experimental features of these phenomena. In particular, in the case of the numerical test reproducing the experiment on the specimen of Lecce stone, the 2D numerical simulation describes, as expected, the formation of concentric rings near the iron source. Moreover, the shape and the intensity of rings depend on the distance from the iron source, as observed experimentally.\\ In the case of the experiment with petri dish filled with agar gel, the numerical simulation is able to reproduce the diffusion of iron from the source located in the inner circle towards the outer circle where another substance is present, and the corresponding formation of iron ring precipitate in the zone of the interface between the two substances is obtained.\\

\paragraph{\bf State of the art and relevant literature.}
In order to explain this phenomenon, many studies have been done both from experimental and from theoretical point of view (see \cite{bams, c, chan, stern, hilhorstetal, mimura} and references therein). The scientific theories developed can be mainly classified into two approaches: pre-nucleation and  post-nucleation.

Pre-nucleation approach is the first attempt to describe the process and states that Liesegang bands form before crystals nucleates, see \cite{Ostwald1}. Among this group, super-saturation effect is the first theory developed and the first one that has been mathematically modeled. In such a family, there is the reaction-diffusion mathematical model introduced by Keller and Rubinow \cite{KR}. In one-dimension, such model is able to reproduce bands which satisfy the spacing law. However, in higher dimensions, the ring patterns obtained numerically do not seem to have radial simmetry. \\
Starting from the 1980's, the post-nucleation approach was developed, see \cite{Kai-Muller}, based upon the Ostwald’s ripening process for colloidal particles, including instability and competitive growth theory. In two space dimensions, the mentioned model, including nucleation effects, does not only show rings formation, but also spiral patterns.\\
A description of the mechanism of dissolution of iron hydroxide and calcite can be found, respectively, in \cite{d} and \cite{e}, while the formation of iron colloidal solution ($Fe^{2+}$ and $Fe^{3+}$) and its chemical reaction in calcareous soil are described, respectively, in \cite{f} and \cite{g}.

\paragraph{\bf Paper organization.}
In Section \ref{sec:labexp} we describe the materials used to reproduce the Liesegangs phenomenon in laboratory, the setting of experiments and the results obtained.
 The mathematical model is described in Section \ref{sec:model}. 
Numerical results obtained with the one-dimensional and two-dimensional model are reported in Section \ref{sec:numtest}. The paper ends up with some conclusions and future perspectives  discussed in Section \ref{sec:concl}.

\section{Materials and methods}\label{sec:labexp}
\begin{figure}
\centering
\includegraphics[height=6 cm, width=7cm]{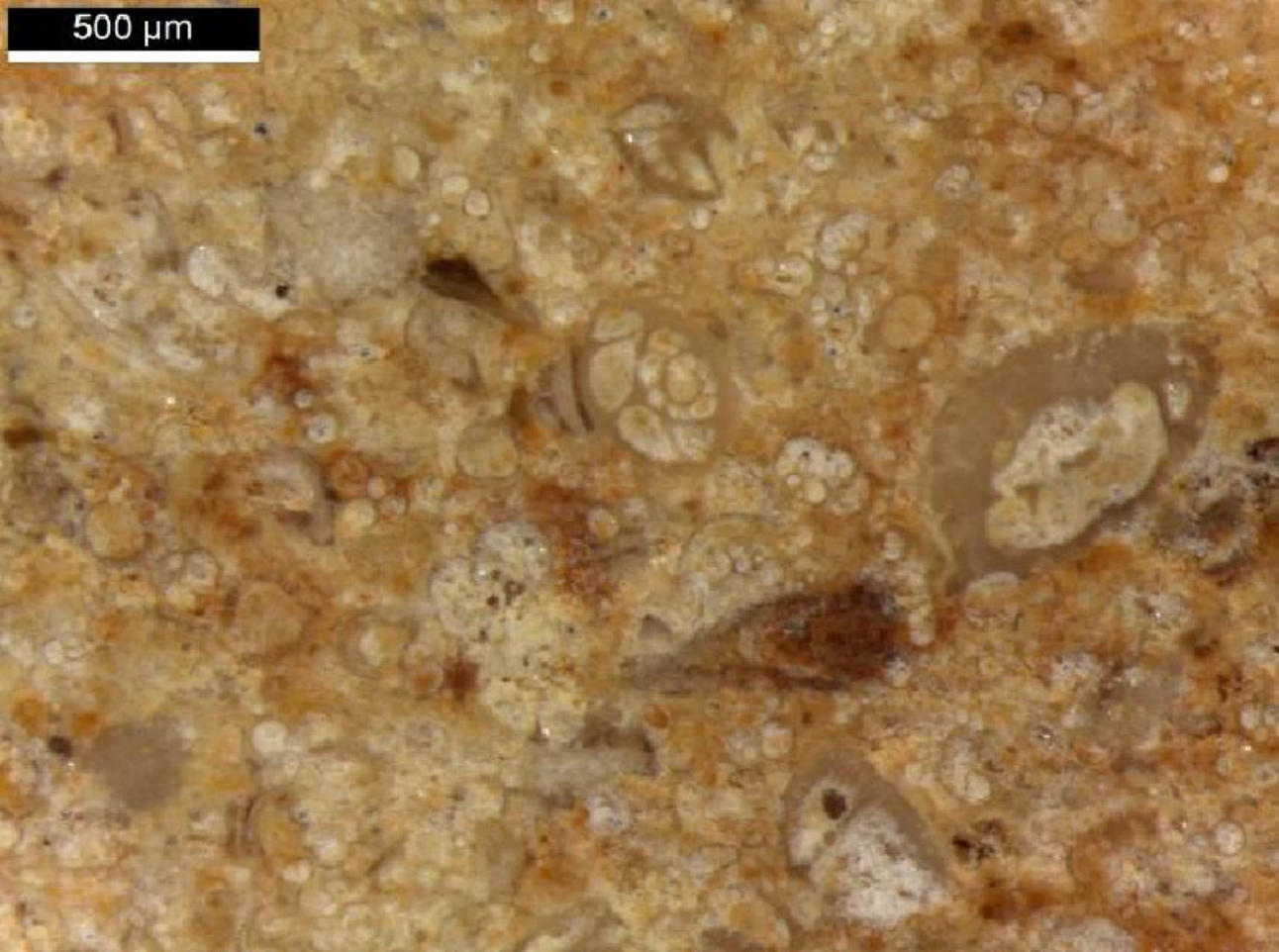}
\caption{Lecce stone, O.M. 50x. Fossils are visible.}
\label{fig:lecce}   
\end{figure}

In order to understand the process of development of rings in a time period shorter respect to the geological one, we decided to reproduce the experiments in laboratory.
Different chemical reactions were used in suitable environments: carbonate stone represented by Lecce stone in the first experiment, and agar gel in the second experiment. 
Lecce stone, coming from the region of Salento in the South of Italy, is a  clay-magnesium limestone, formed by sands of calcareous rocks and elements of organic origin, such as fragment of corals, mammals skeletons and various microscopic sea fossils immersed in calcite cement \cite{a}.
With the petrographic exam, the Lecce stone appears to be composed by a granular mixture inglobated in the calcite cement, thus the main constituent is calcium carbonate. Its structure is porous, fine, homogeneous and gold yellow grain.
Besides the calcareous elements, one can see grains and fragments of glauconia (iron and magnesium  silicate), a yellow substance derived principally from organic leftovers, very small grains of iron oxide and a variable quantity of silica, partly deriving from the degradation of  glauconia (see Figure \ref{fig:lecce}).

The studies in literature regarding the porosity of the Lecce stone \cite{falchi} via mercury porosimetry, show a high open porosity (approximately of 40$\%$) and a bimodal distribution of diameter of pores with maxima of about $0.5$ and $5 \ \mu m$.

\subsection{Experiment I: iron precipitated on Lecce stone.}\label{exp1}

\begin{figure}
\centering
\includegraphics[height=7 cm, width=7cm]{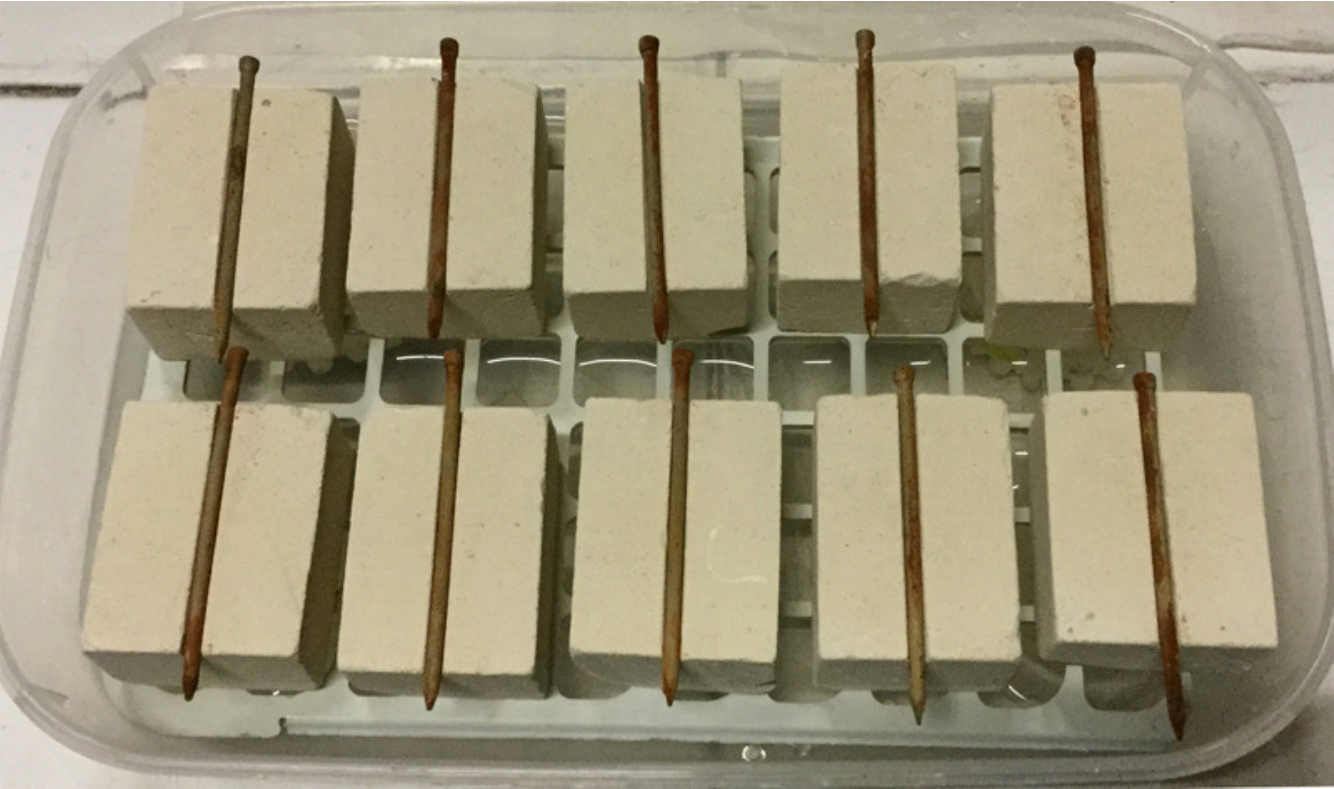}
\caption{Setting and placing of Lecce stone specimens in the bucket.}  \label{fig:allest}   
\end{figure}

As preliminary tests, imbibition experiments on Lecce stone were conducted, according to standard UNI 10859 Protocol "Cultural Heritage - Natural and artificial stones" (determination of water absorption by capillarity) and Normal 7/81 Protocol Cultural Heritage - Natural and artificial stones" (imbibition capability). It served as a control sample to test transport properties of the materials under study. The absorption curves obtained show that saturation of Lecce stone is reached rapidly (few minutes), for this reason, in the modelization of Liesegang phenomenon, in this framework we are neglecting the effects due to the presence of water flowing in the stone, see Section \ref{sec:model}. \\
In order to reproduce Liesegang rings is necessary to have as starting point a source of iron ions and a "host" stone. We decided to use  Lecce stone for its high permeability, porosity and homogeneity, since such features allow a rapid reproduction of the phenomen of Liesegang rings.\\ 
The reactant used consists of element of pure iron,  oxidized in the presence of humidity, and it provided the result of the formation of colloidal gel containing ions $Fe^{2+}$, that successively precipitate as oxide of hydrate iron $Fe$-$(OH)_2$, which are able to migrate in a porous medium.
The tests on the growth of Liesegang rings  have been conducted using 10 blocks of Lecce stone of dimension $30 \times 50 \times  25 \ mm^3$, suitably fluted along the middle of the superior side to host the iron element.
Then, the blocks have been placed on a horizontal grid suspended in a PVC bucket containing at the bottom 500 ml of $H_2O$ demineralized, see Figure \ref{fig:allest}. The bucket was successively closed with a perforated cover to allow air circulation, and placed in an incubator at  the temperature of $30^\circ$C; under these conditions it was reached a humidity  of 80$\%$ after 24 hours.

\begin{figure}
\includegraphics[width=7cm, height=7cm]{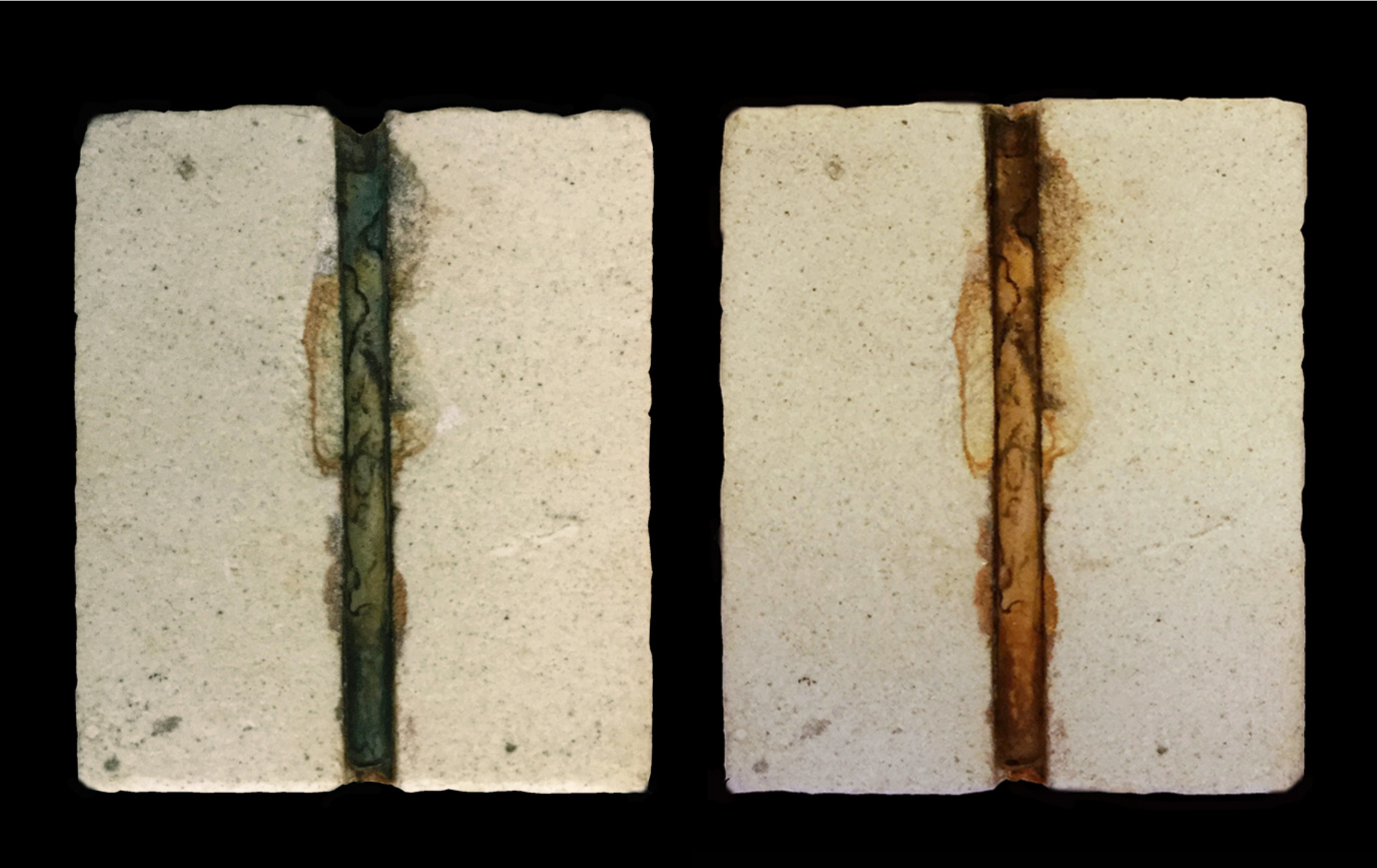}
\caption{Setting of the iron nail: in the specimen on the left it is evident the green color of the GR, turning to ochre in the specimen on the right.}\label{figura13}
\end{figure}

The monitoring of the Lecce stone blocks was every 3 days. The oxydation phenomenon of nails and the consequent formation of colloidal iron solution was slow. 
We observed a consistent corrosion product after 15 days only on a specimen.
Therefore, we decided to replace the nails with common carpentry iron wire with a section of $3 \ mm$. In such a way, we could see results after 5 days. We also noticed that the formation of oxydes is not homogeneous. 
Moreover,  the part of the nail in contact with the carbonate matrix results to be green (GR) but it fastly turns to ochre in contact with the air, see Figure \ref{figura13}.
 
With a visual monitoring of the oxidation process, we decided to extract a single block from the bucket every 5 days and to cut it with a section of about 3.5 mm of thickness. 
Due to inhomogeneities, both in the time evolution and the concentration of the corrosion product, we decided to insert again each specimen, already extracted and sectioned, in the bucket, in order to observe  the kinetic of rings development on the same specimen.

\begin{figure}
\centering
\includegraphics[height=8 cm, width=6cm]{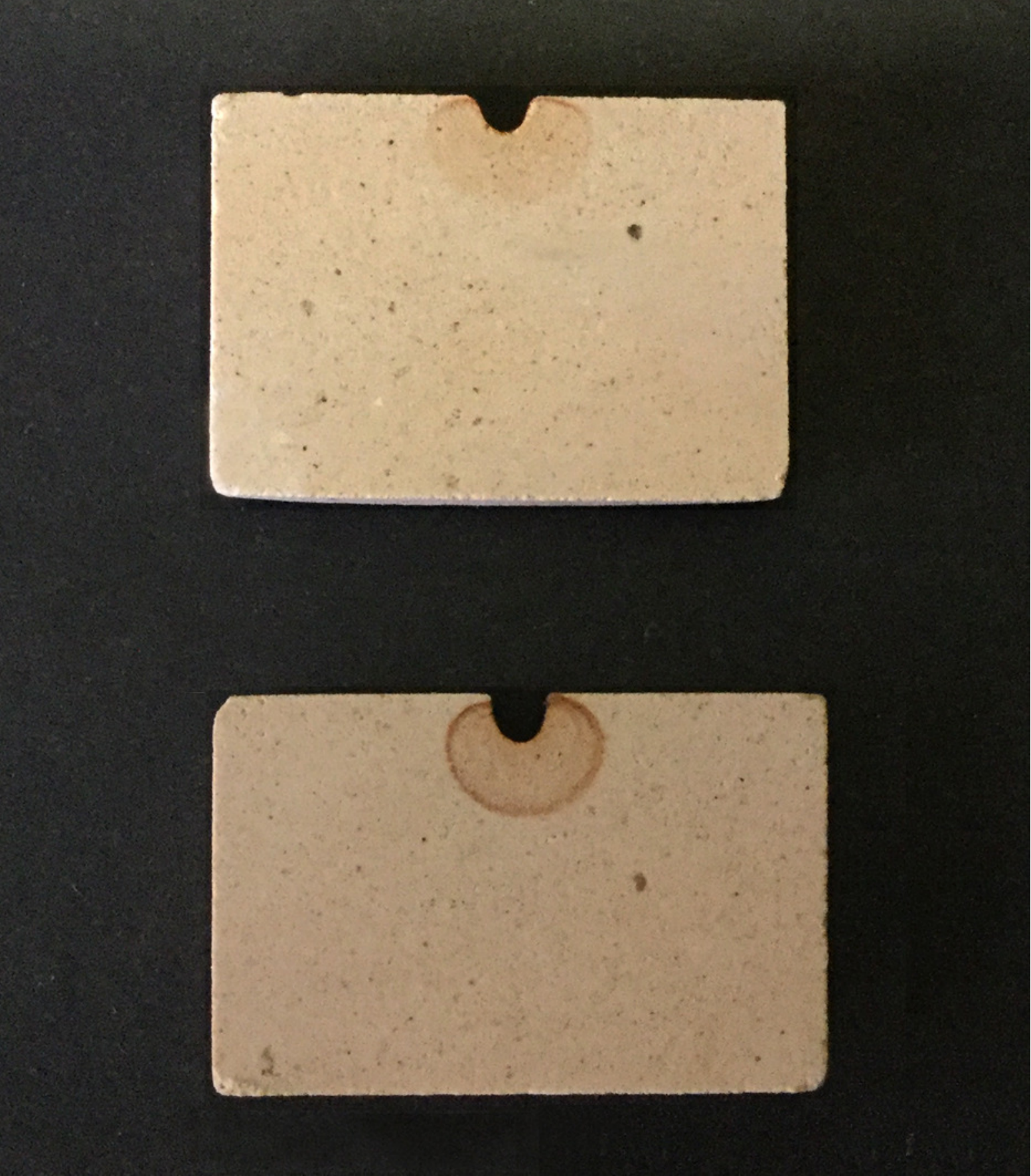}
\caption{Iron rings on a specimen sections after 5 days (top) and after 15 days (bottom) of exposure.}  \label{fig:provini}   
\end{figure}

In Figure \ref{fig:provini}, we can observe the first two sections of the specimen extracted after, respectively, 5 days (top) and 15 days (bottom) of exposure: both the sections present the rings formation.
In particular, after 5 days (see Figure \ref{fig:provini} a) we notice the formation of a slight circular area, with dimensions which are comparable to that obtained after 15 days (see  Figure \ref{fig:provini} b). The ring in the section in Figure \ref{fig:provini} (b) has the area of diffusion homogeneous, well defined, with a higher concentration of iron on the boundary. We also notice that the bands of the ring, in both the sections,  are more marked in the upper part, and this is probably due to the proximity of the iron source  determining a greater accumulation of corrosion product.\\  

\begin{figure}
\centering
\includegraphics[height=6 cm, width=6cm]{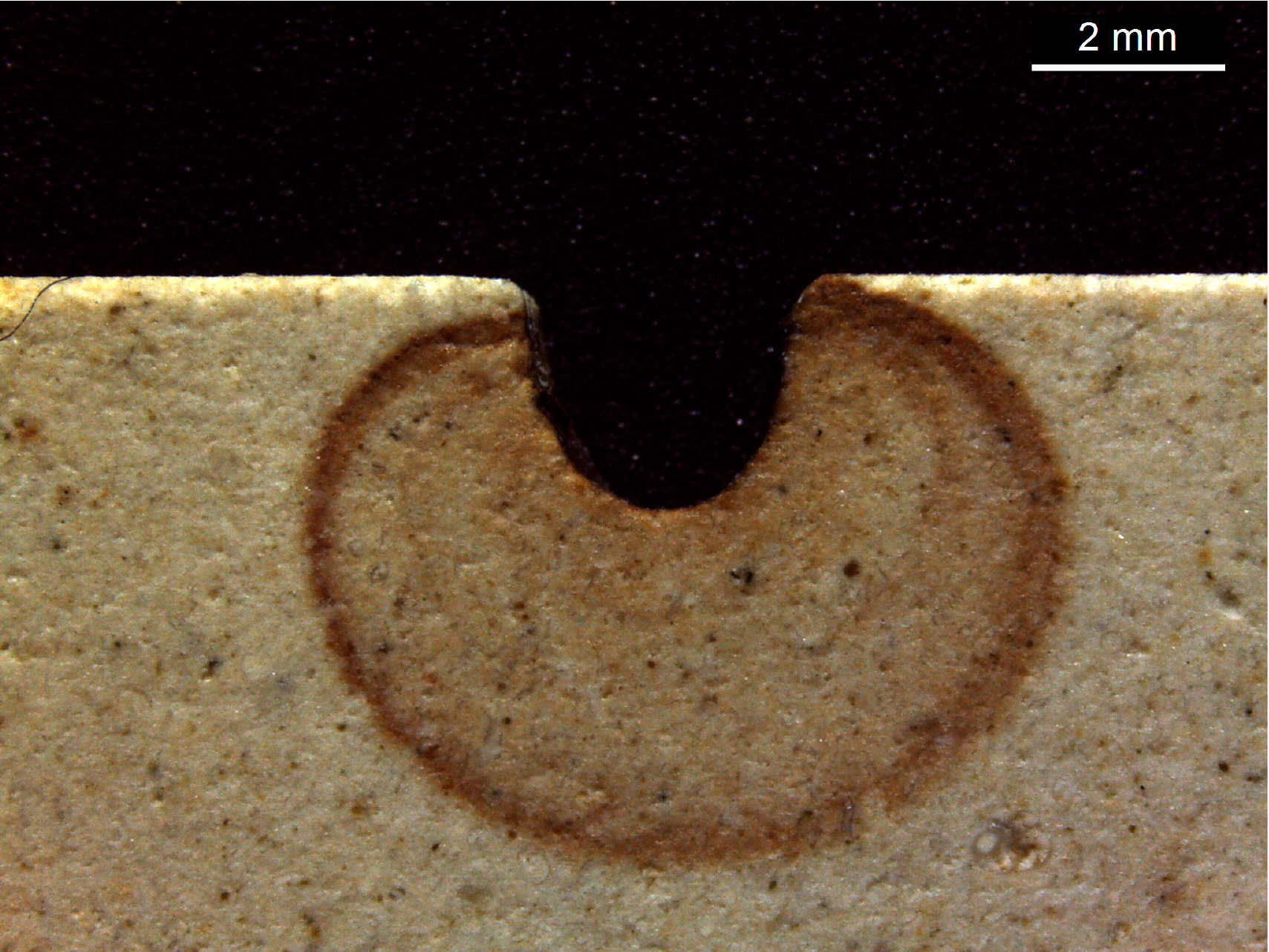}\quad \includegraphics[height=6 cm, width=6cm]{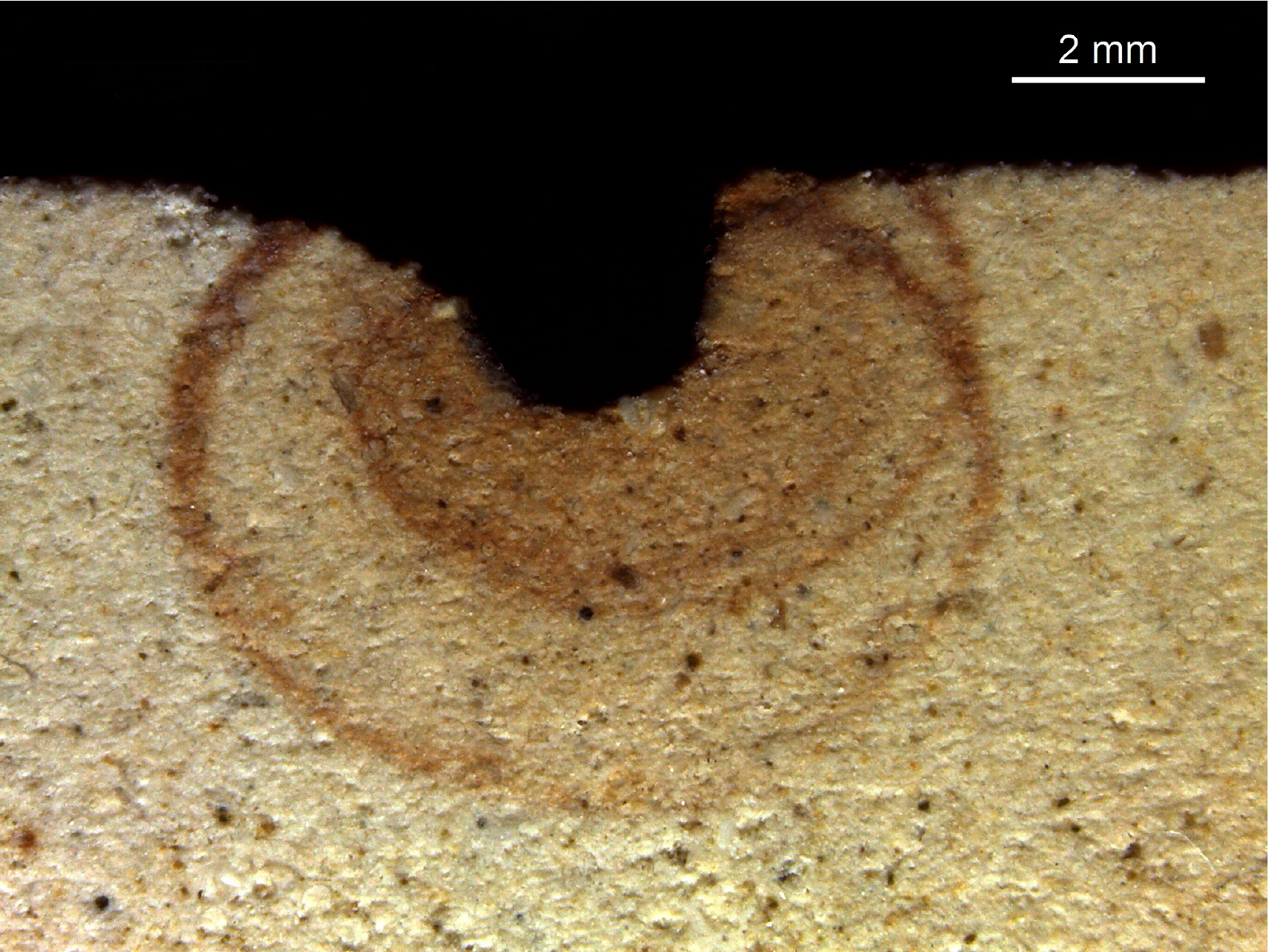}
\caption{Iron rings on the specimen sections after 14 days (left) and after 27 days (right) of exposure.}  \label{fig:rita}   
\end{figure}

\begin{figure}
\centering
\includegraphics[height=7 cm, width=7cm]{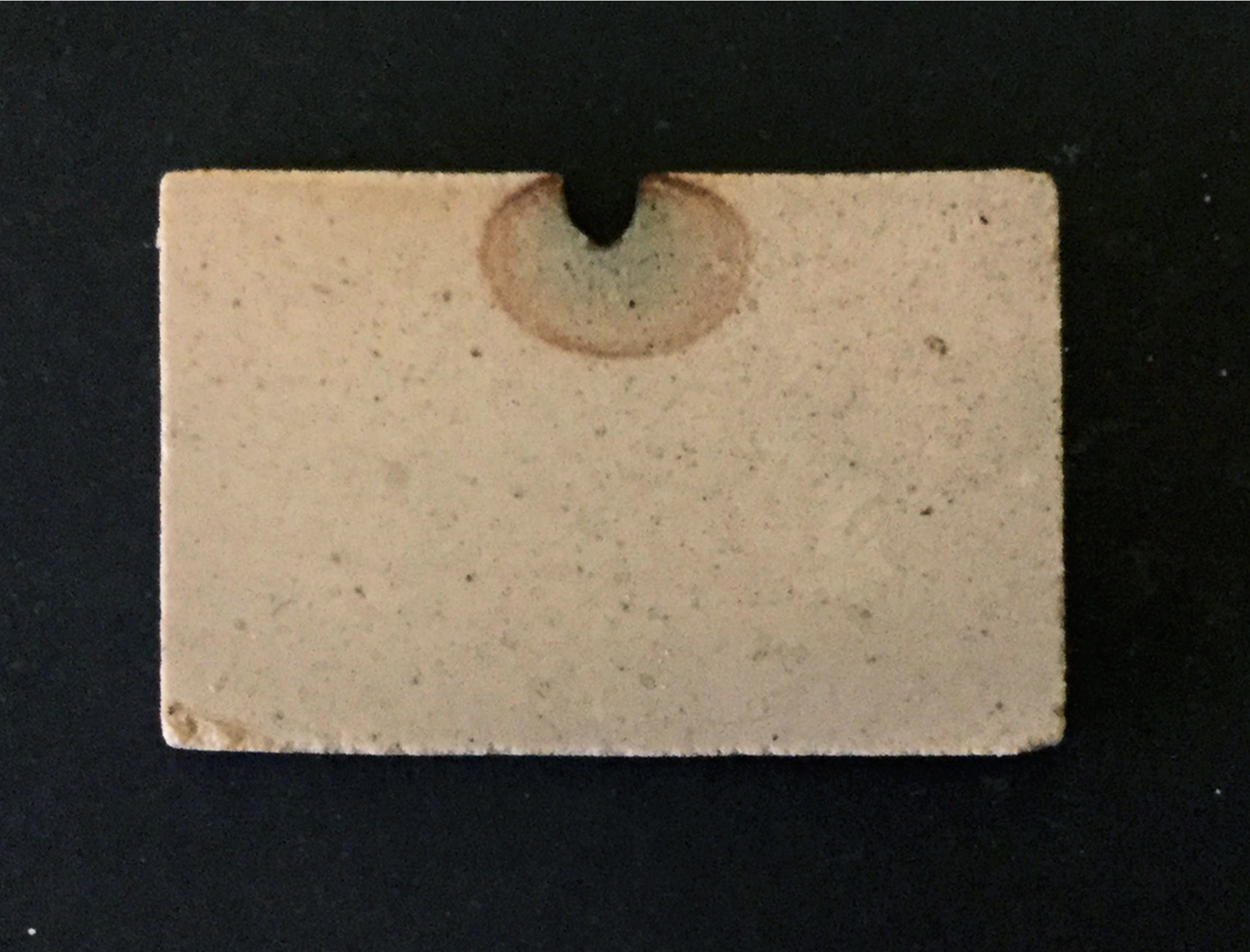}
\caption{Image of green precipitated in the iron rings of the inner cut section.}  \label{fig:green}   
\end{figure}

We remark that in the images above, the portion of the block interested by the phenomenon of Liesegang rings is small and is about $13$ $\times$ $8 \ mm^2$, thus we will focus on this small area to conduct numerical simulations presented in next Section \ref{sec:numtest}.

Another interesting aspect is the presence of a precipitated of green color in the inner cut section  (see Figure \ref{fig:green}).
The green area can be addressed to the presence of "green rust" ($Fe^{2+}_3 Fe^{3+}_2 (OH^-)_{12} \cdot CO_3^{2-}\cdot H_2O$ type carbonate I), an unstable product of iron corrosion, that disappears for oxidation in the air \cite{b, 3}.
From the sections obtained, we acquired the images with microscope  Leica M205, with 10x zooms, and successively elaborated with the software for image analysis "ImageJ".

\subsection{Experiment II: iron precipitated in agar gel}\label{agar-agar}

The agar gel for the study of the precipitation reactions has two advantages: is a porous medium that allows the diffusive transport of small molecules and the visual monitoring during and after the reactions it is easy, \cite{chan}. 
In the present experiment, the gels realized are solutions with both the reactants $FeSO_4$ and $NaOH$, with molar concentration, respectively, 0.1 M and 0.2 M, determining as reaction product the Green Rust formation \cite{3}.
With each of the two reactants at molar concentration 0.2 M, two solutions are prepared with 4$\%$ of agar. Then, each solution is spilled into a petri dish of 60 mm of diameter, to obtain two discs cutted inside with a hole of 20 mm and into the holes the agar gel formed with the opposite reactant is inserted. In such a way, we obtain two systems: the first one constituted by agar-$FeSO_4$ (with 0.2 M and pH $\sim$ 5) at the exterior and agar-$NaOH$ (with 0.1 M and pH 12.3) at the interior, and the second one by agar-$NaOH$ (with 0.2 M and pH $\sim$ 12.0) at the exterior and agar-$FeSO_4$ (with 0.1 M and pH $\sim$ 5.4) at the interior. 

\begin{figure}
\centering
\includegraphics[height=6 cm, width=6cm]{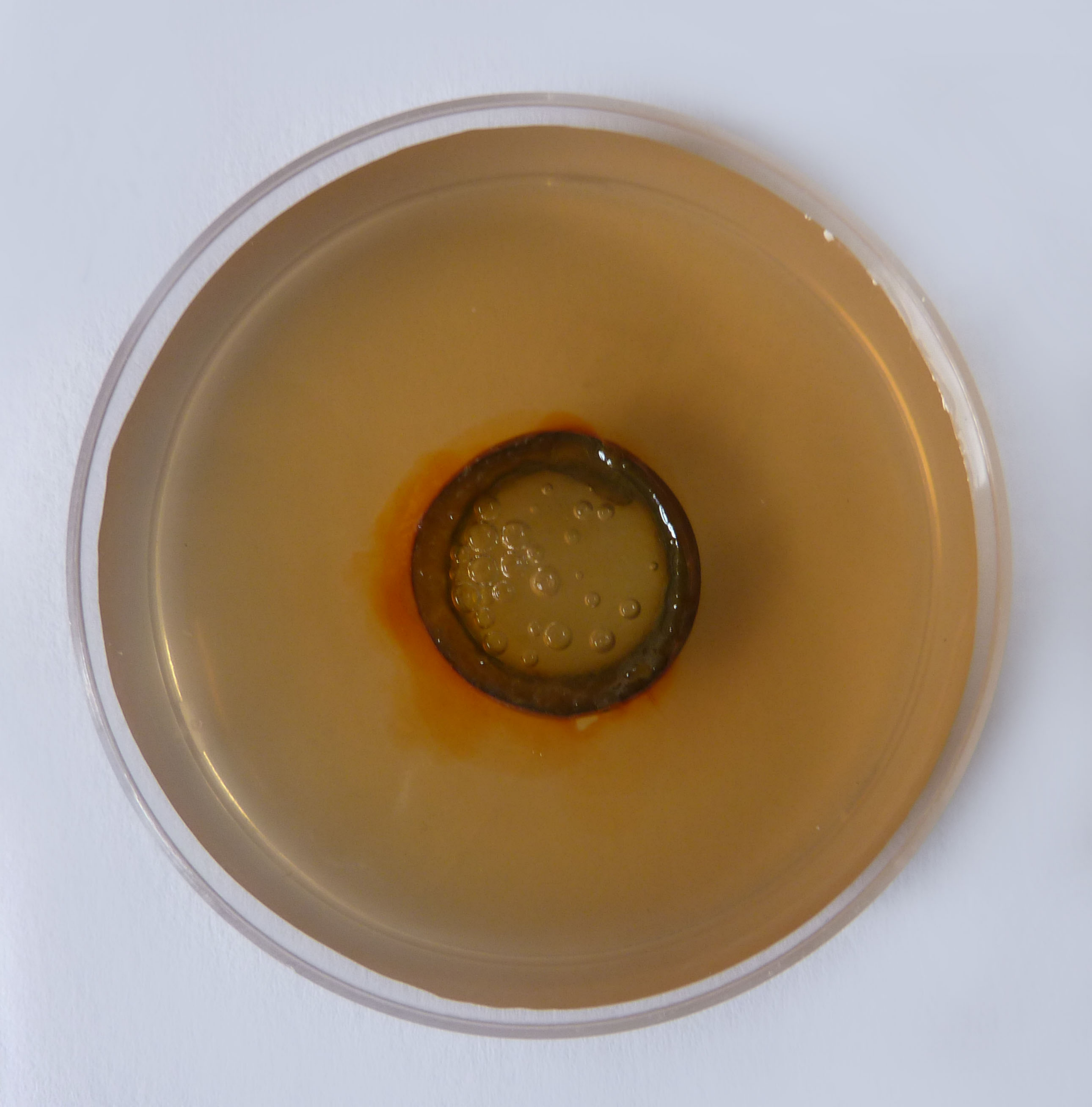}\quad
\includegraphics[height=6 cm, width=6cm]{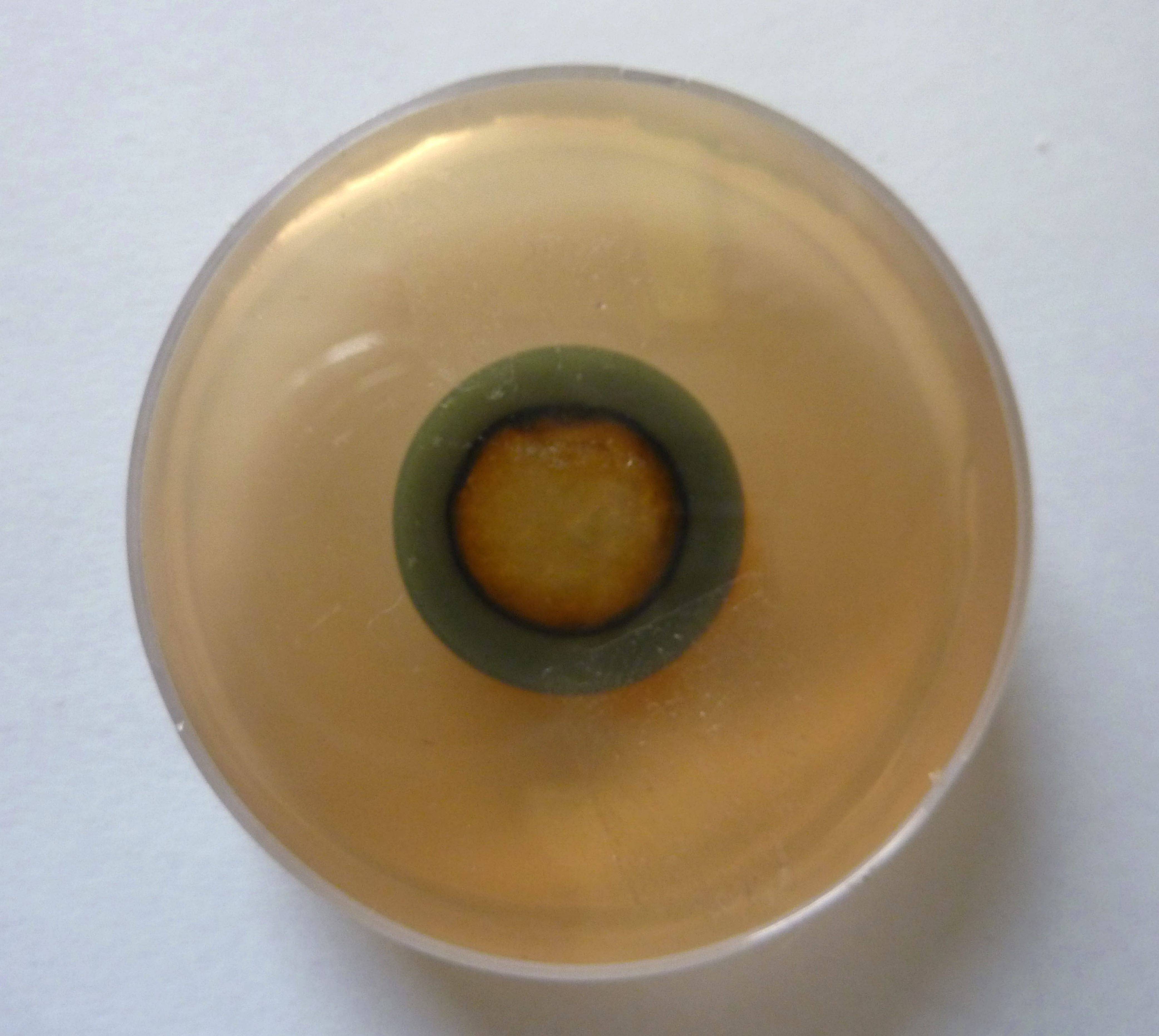}
\caption{Front (on the left) and back (on the right) of the agar dish after 8 hours.}  \label{fig:agar}   
\end{figure}

After 8 hours, in the system $NaOH$ outside and $FeSO_4$ inside, the reaction produces, on the visible part of the dish, a ring of color brown-red diffusing from the center towards the gel containing $NaOH$ (see the left image in Figure \ref{fig:agar}). Looking at the same system from the back, two concentric rings can be observed: the inner one is thinner, of color blue-dark green, with thickness of about 0.8 mm, and the outer one of color  blue-light green, with thickness of about 4 mm (see the right image in Figure \ref{fig:agar})).
We can suppose that the blue-green color due to the precipitation of GR-$SO_4^2$, is visible only on the back of the disc for a minor concentration of oxygen, since the lack of oxygen determines a slower kinetic of the oxydation of anion $Fe^{2+}$ to $Fe^{3+}$.\\
The results is slightly different if we invert the order of reactants.
In the system  $FeSO_4$ outside and $NaOH$ inside, after 8 hours, differently from the system described above, both in the visible part and in the back of the capsule, a green ring with thickness of about 1.2 mm is formed, while at the interior the gel is of red-brown color.\\
In both the tests, after 24 hours, the solution contained in the inner ring is dissolved. 
The solution containing iron, in the presence of a gradient, thus seems to diffuse towards the zone with lower concentration.

\section{Mathematical modeling of iron rings}\label{sec:model}

With the aim of reproducing the iron rings on Lecce stone shown in Section \ref{sec:labexp}, we considered and compared the results obtained with different mathematical models, see for instance \cite{Kai}, \cite{KR}, \cite{mimura}. Analyzing the results obtained, we decided to use the one-dimensional and two-dimensional Keller-Rubinow  model and a modification of it, as we will explain in the sequel. The modification is due to instability problems in the original model.\\ 
Note that at the moment we are neglecting the effects due to the presence of water flowing in the stone, since the absorption curves found experimentally suggest a fast reach of the asymptotic condition of saturation (few minutes). Indeed, this is a very short time compared to the time needed to have the development of Liesegang rings, which is of order of days or weeks.

 \subsection{KR model}
The classical KR model reproduces the reaction of $a$, the outer electrolite, with the inner electrolite $b$, while $c$ is the reaction product and $d$ is the precipitate of $c$.
The KR model is described by the following reaction-diffusion system:
\begin{eqnarray}\label{KR}
  \left\{\begin{array}{lll}
\frac{\partial}{\partial t}  a &=& D_a \Delta a - k a b,  \\
 \frac{\partial}{\partial t} b &=& D_b \Delta b - k a b, \\
\frac{\partial}{\partial t} c &=& D_c \Delta c + k a b - q  P(c,d), \\
\frac{\partial}{\partial t} d &=& q  P(c,d),
  \end{array}\right.
\end{eqnarray}
where $a,b,c,d$ are the densities of each substance, $D_a, D_b, D_c$ are the diffusion coefficients, $k$ is the chemical reaction constant, $q$ is the precipitation coefficient and $P(c,d)$ the precipitation term as defined in the Keller-Rubinow  model:
\begin{equation}\label{P}
P(c,d)=\left\{
\begin{array}{ccc}
0&\text{ if }&c<C^*\text{ and }d=0\\
(c-C_s)_+&\text{ if }&c\geq C^*\text{ or }d>0,
\end{array}\right.
\end{equation}
where $C_s$ is the saturation density of $c$, $C^*$ is the super-saturation density of $c$, $C^*>C_s$, and function $(\cdot)_+$ is the function positive part.
Following the experimental settings we have that the solid matrix can be constituted by calcium carbonate (if the host is Lecce stone - experiment I) or a colloidal solution with sodium hydroxide (if the host is agar gel in petri dish - experiment II). \\
Let us detail the variable that come into play in the two cases:

\begin{enumerate}
\item $a$ in KR is:
\begin{itemize}
\item a colloidal solution composed by the two species of iron: \\
$6Fe(II,III) + O_2 + H_2O$ (experiment I); 
\item a colloidal solution composed by ferrous sulphate $FeSO_4$ (experiment II);
\end{itemize}
 \item $b$ in KR can be: 
 \begin{itemize}
 \item carbonate ion/calcite: $CO_3^{2-}$ (experiment I);
 \item sodium hydroxide $NaOH$ (experiment II);
 \end{itemize}
\item $c$ in KR  is the the reaction product: green rust;
\item $d$ in KR  is the precipitate of $c$: ex-green rust.
\end{enumerate}

The phenomenon of Liesegang rings formation in the stone can be explained as follows. Substance $a$ diffuses into the host stone and reacts with $b$ to produce green-rust $c$, composed by layers of $Fe^{2+}$ and $Fe^{3+}$. Successively, the intermediate product green rust,  precipitate as a solid $d$. The conditions determining the precipitation and ossidation of green rust are supposed to happen when the concentration of the iron species exceedes the solubility limit. At enviromental temperature and inferior, the ex-green rust is stable even after long time. 
The precipitation of a new band starts when the local concentrations overcome again the solubility product. Then, the critical factor is represented by the ratio between the two diffusion velocity of $a$ and $b$.

\section{Numerical tests}\label{sec:numtest}
We tested the 1D and 2D the KR model and we found a quite strong numerical instability of the solution. In particular, observing the precipitate substance $d$, we noticed that Liesegang bands may vary in width, place and intensity, depending on the space grid discretization as shown in the next Section \ref{test_KR1D}. 


\subsection{Preliminary test with KR model in one space dimension}\label{test_KR1D}
For the 1D case, the initial and boundary conditions assumed for the KR model (\ref{KR}) are:
\begin{equation}\label{IB_conditions}
 \left\{\begin{array}{ll}
a(x,0)= 0, \ b(x,0)=b_0, c(x,0)=d(x,0)=0, \ x \in [0,L],\\
 a_x(0,t)=a_0, b_x(0,t)=c_x(0,t)=0, \ t>0,\\
a_x(L,t)=b_x(L,t)= c_x(L,t)= 0, \ t>0,
 \end{array}\right.
\end{equation}
with $a_0, b_0 \in \mathcal{R}^+$. \\
In particular, following \cite{mimura}, parameters of model (\ref{KR}) are assumed as reported in Table \ref{table:param}.
\begin{table}[h]
\vspace{0.5cm}
\center
\begin{tabular}{l c l }
\hline
$a_0$ & = & 10\\
$b_0$ & = & 1\\
$D_a $  &  =  & 1e-3  \\ 
$D_b $  &  =  & 1e-3\\ 
$D_c $  &  =  & 1e-3 \\ 
$k$  &  =  &  50  \\ 
$q$  &  =  &  50 \\ 
$ C_{s} $  &  =  & 0.2 \\ 
$ C^{*} $  &  =  & 0.8 \\ 
$L$ &  =  & 1.5  \\ 
\hline
\end{tabular}
\vspace{0.5cm}
\caption{Parameters of the 1D simulations of the model (\ref{KR}).}
\label{table:param}
\end{table}

\vspace{0.2cm}
\paragraph{\bf Numerical approximation of 1D system (\ref{KR}).}
We mesh the interval $[0,L]$ with a step $\Delta x = \frac{L}{N}$ and we denote 
$$
\lambda = \frac{\Delta t}{\Delta x}, x_j = j \Delta x, j=1,...,N. 
$$
We also set $w^n_j=w(x_j,t_n)$ the approximation of the function $w$ at the height $x_j$ and at the time $t_n$. 

Following \cite{BGIMN}, the first three equations in (\ref{KR}) are numerically solved by finite difference schemes. Such approximation consists of a Crank-Nicolson solution method in time,  second order central finite difference in space, combined with a nonstandard approximation for source terms.\\
In particular, the evolution of the concentrations of problem (\ref{KR}) is computed by solving the following system composed by coupled equations:

\begin{equation}\label{KR_discr:1}
 \left\{\begin{array}{ll}
\frac{a^{n+1}_j - a^n_j}{\Delta t} = D_a \frac{\delta^2 a^{n+1} + \delta^2 a^{n}}{2} - \frac{k}{2} (a_j^{n+1}b_j^n + a_j^n b_j^{n+1}),\\
\frac{b^{n+1}_j - b^n_j}{\Delta t} = D_b \frac{\delta^2 b^{n+1} + \delta^2 b^{n}}{2} - \frac{k}{2} (b_j^{n+1}a_j^n + b_j^n a_j^{n+1}),\\
\frac{c^{n+1}_j - c^n_j}{\Delta t} = D_c \frac{\delta^2 c^{n+1} + \delta^2 c^{n}}{2} + \frac{k}{2} (a_j^{n+1}b_j^n + a_j^n b_j^{n+1})- q P(c_j^n,d_j^n),\\
\frac{d^{n+1}_j - d^n_j}{\Delta t} = q P(c_j^n,d_j^n), 
 \end{array}\right.
\end{equation}
where $\delta^2$ represents:
\be
\delta^2 w = \frac{w^n_{j+1} - 2w^n_j + w^n_{j-1}}{\Delta x^2},
\ee
with $w$ the function to be approximated.
Let us rewrite (\ref{KR_discr:1}) in the form more suitable for implementation:
\begin{equation}\label{KR_discr}
 \left\{\begin{array}{ll}
a^{n+1}_j  = a^n_j + \Delta t D_a \frac{\delta^2 a^{n+1} + \delta^2 a^{n}}{2} - \Delta t \frac{k}{2} (a_j^{n+1}b_j^n + a_j^n b_j^{n+1}),\\
 b^{n+1}_j = b^n_j + \Delta t D_b \frac{\delta^2 b^{n+1} + \delta^2 b^{n}}{2} - \Delta t \frac{k}{2} (b_j^{n+1}a_j^n + b_j^n a_j^{n+1}),\\
c^{n+1}_j = c^n_j + \Delta t D_c \frac{\delta^2 c^{n+1} + \delta^2 c^{n}}{2} + \Delta t \frac{k}{2} (a_j^{n+1}b_j^n + a_j^n b_j^{n+1})- \Delta t q P(c_j^n,d_j^n),\\
d^{n+1}_j = d^n_j + \Delta t q P(c_j^n,d_j^n).
 \end{array}\right.
\end{equation}
The adopted source terms approximation ensures numerical stability and avoids high computational cost while preserving the total mass.
At each time step, the nonsymmetric linear system obtained by the space approximation is solved by the biconjugate gradient stabilized method (BiCGSTAB). 

\begin{figure}
\centering\includegraphics[scale=0.2]{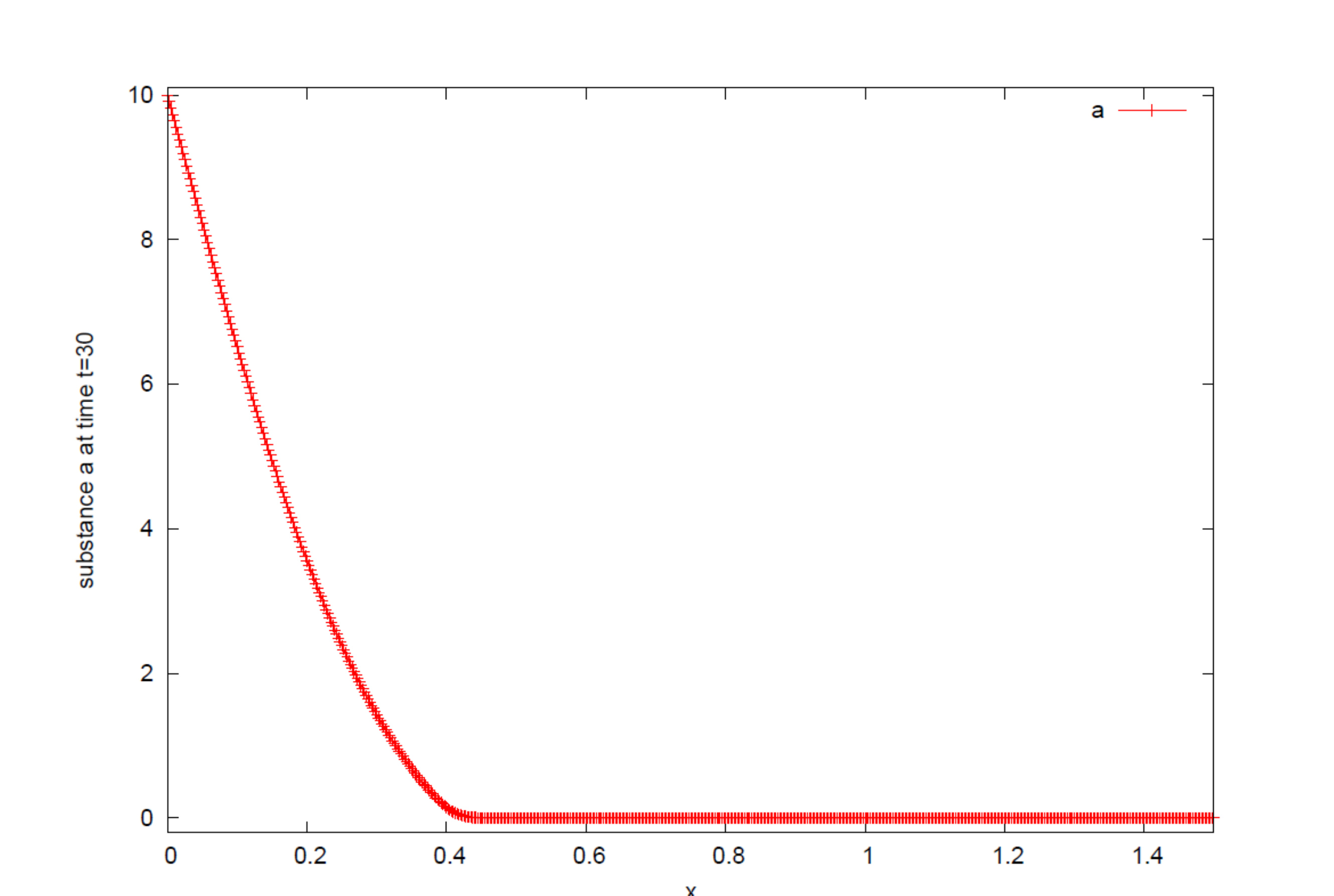}\quad \includegraphics[scale=0.2]{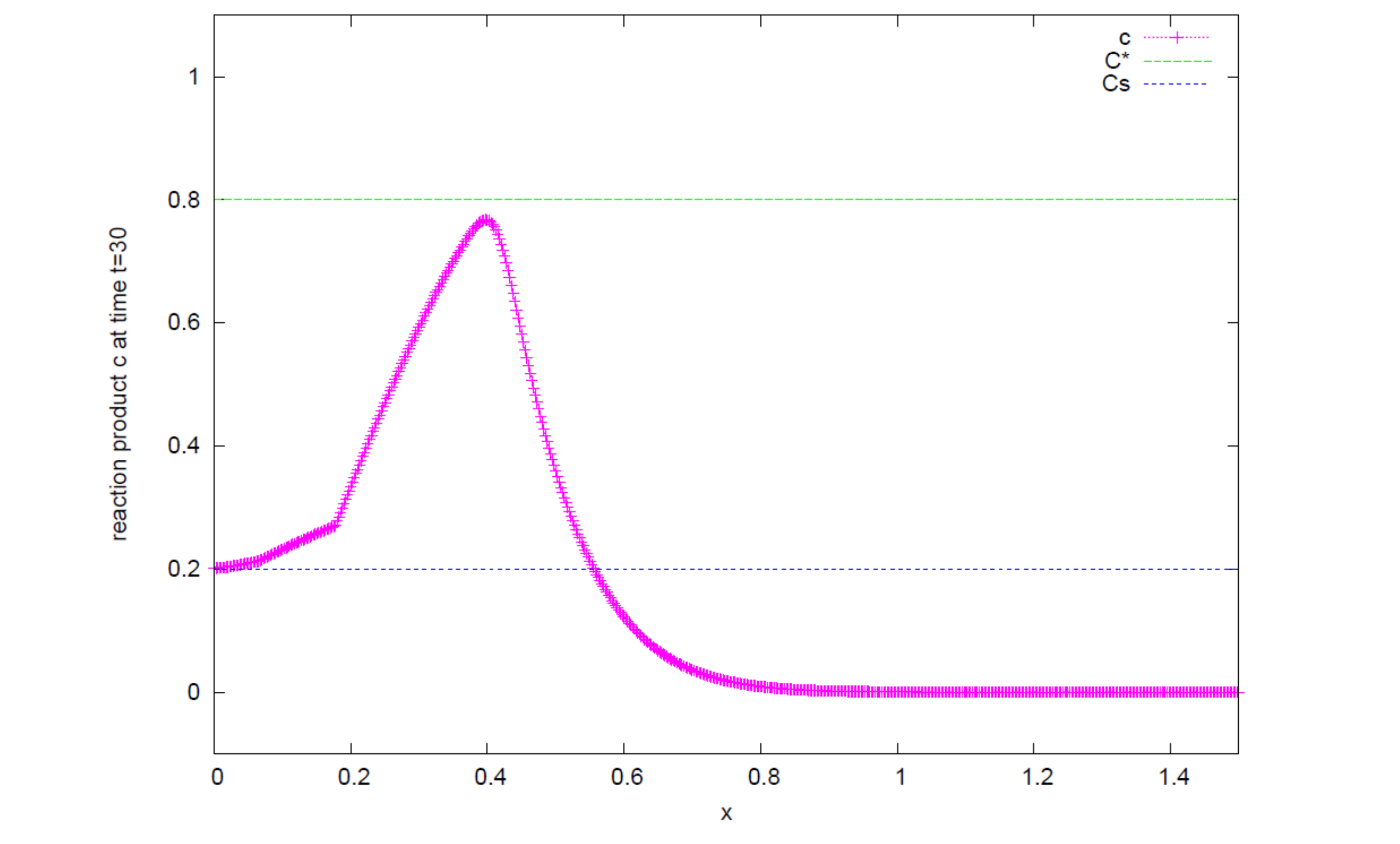}\\
\includegraphics[scale=0.2]{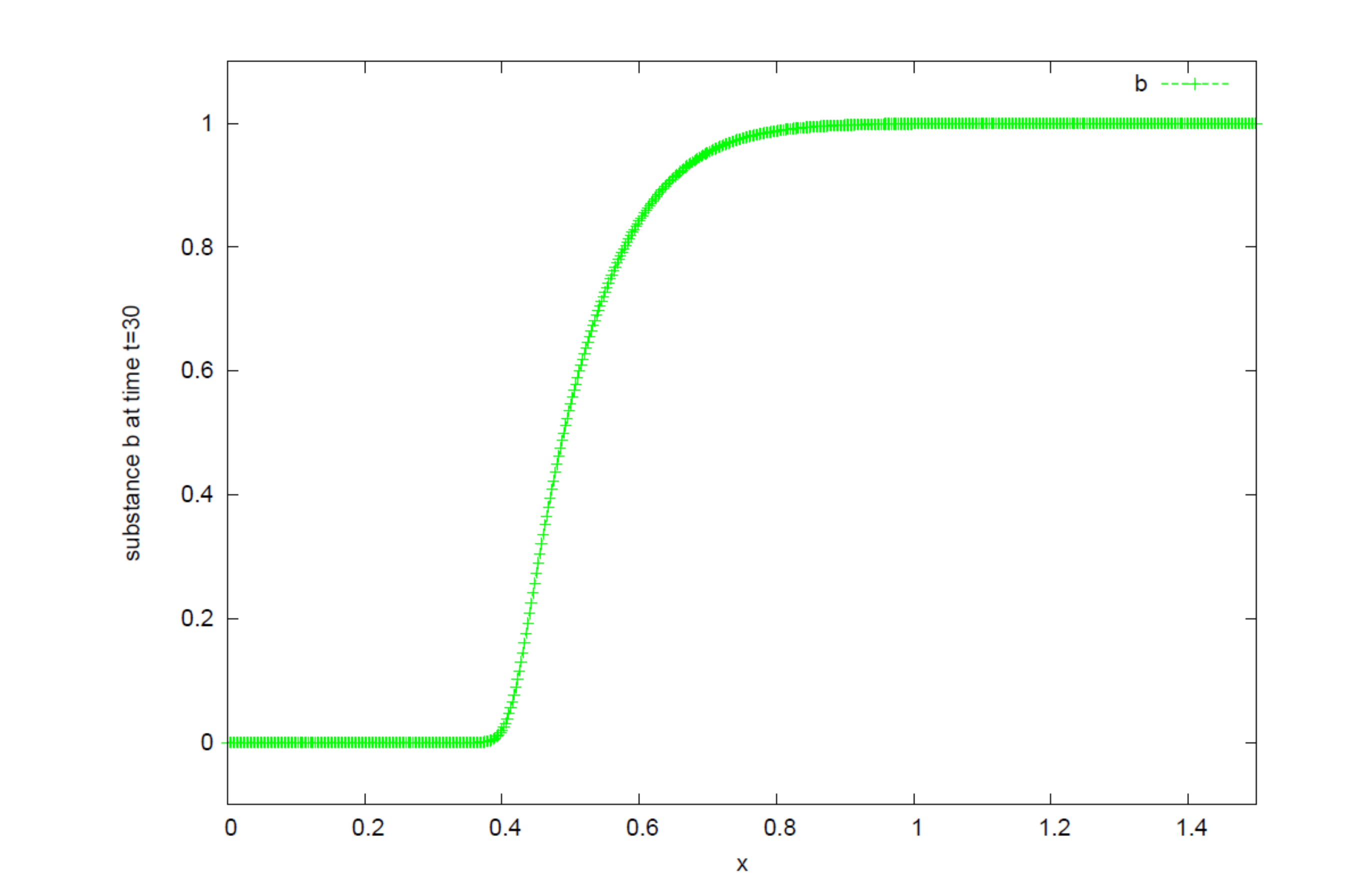}\quad\includegraphics[scale=0.2]{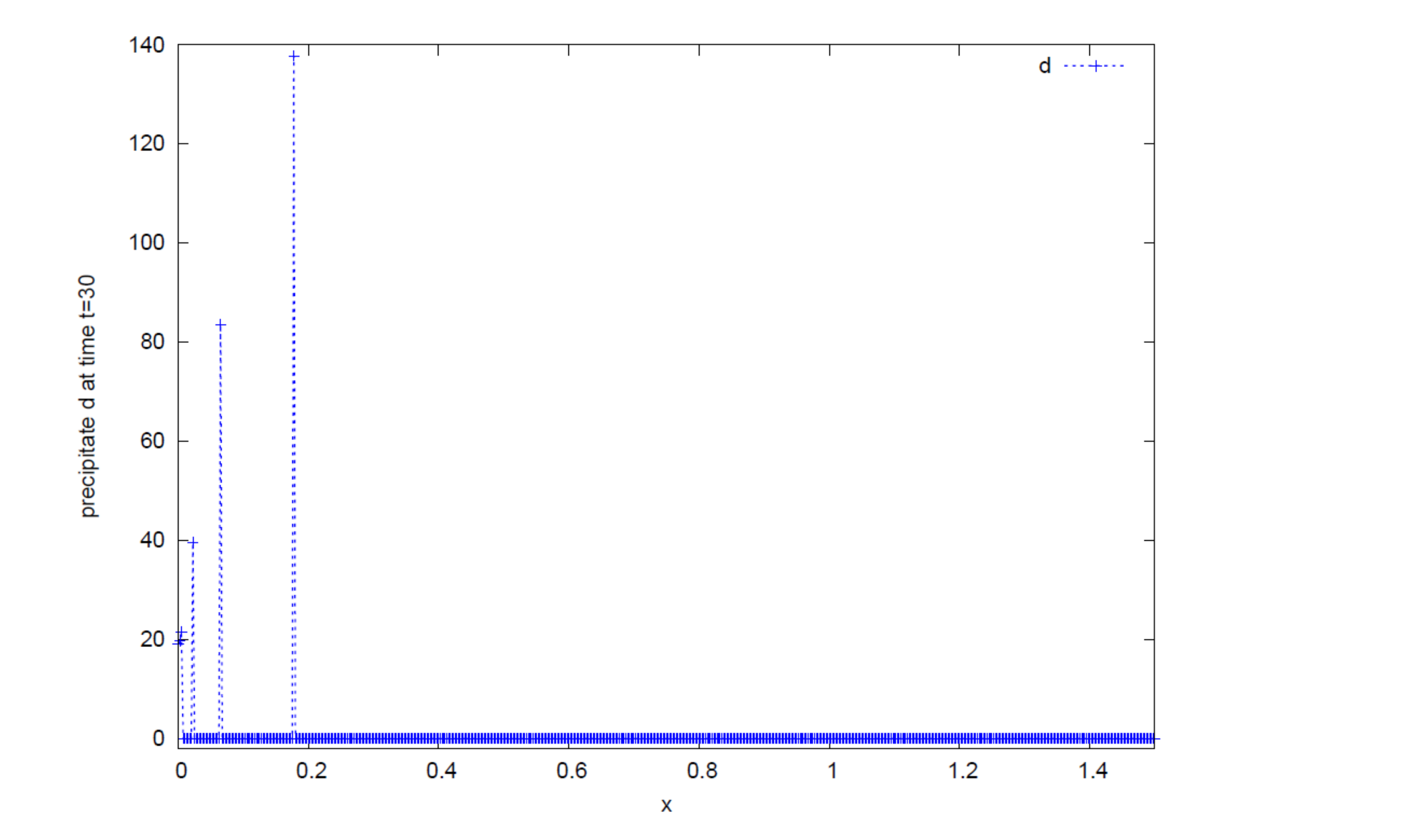}\\
\caption{Chemical substances for the model (\ref{KR}) at time $t=30$, for $\Delta x=0.0025$ and $\Delta t=0.00625$. }  \label{fig:1}   
\end{figure}

The results obtained using initial and boundary conditions above are reported in the graphs of Figure \ref{fig:1}; we can observe the formation of Liesegang bands in the graph of the precipitate substance $d$ (bottom-right graph).\\

\begin{figure}[h]
\centering
\includegraphics[width=3.4cm]{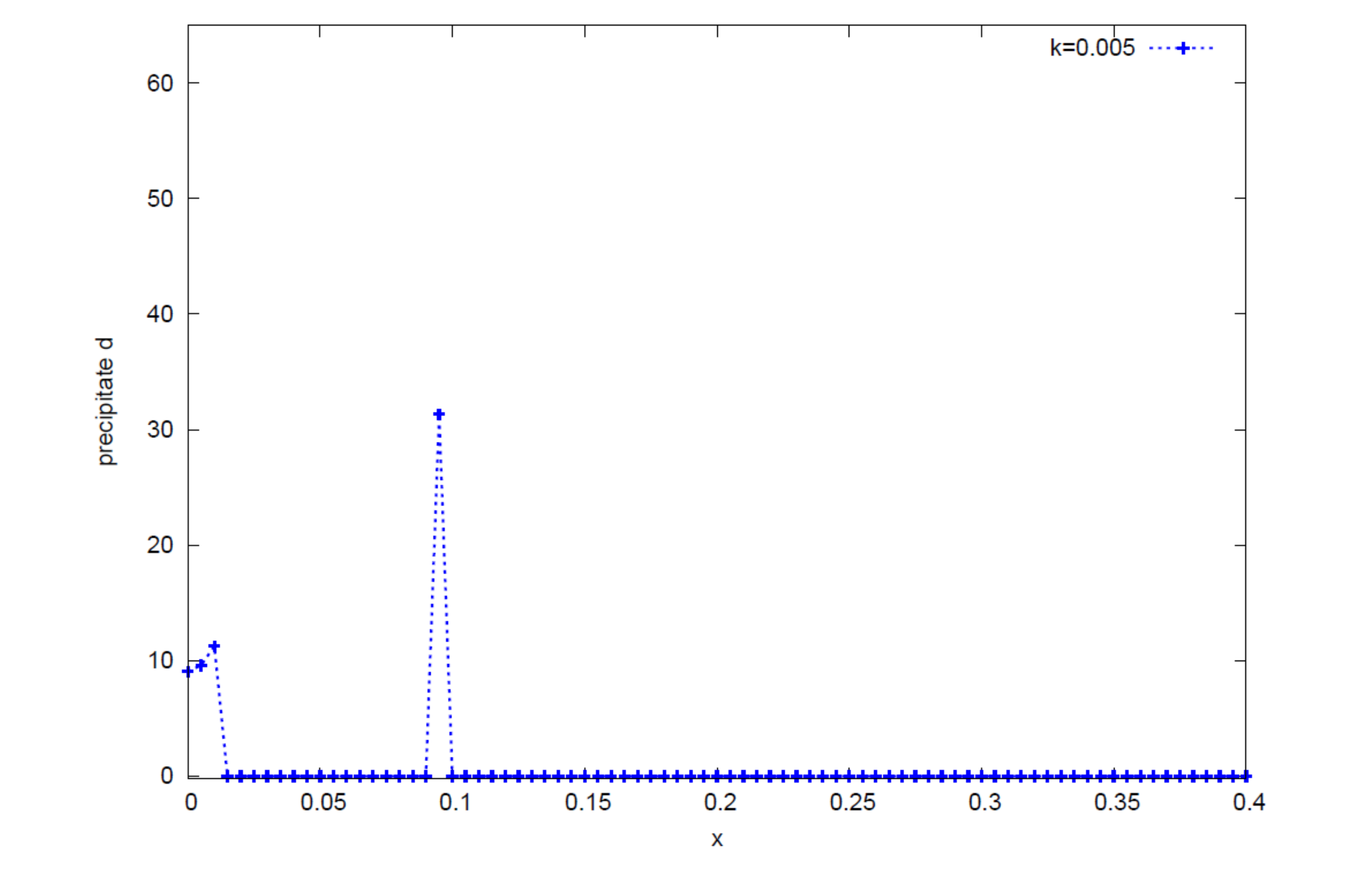}\quad
\includegraphics[width=3.8cm]{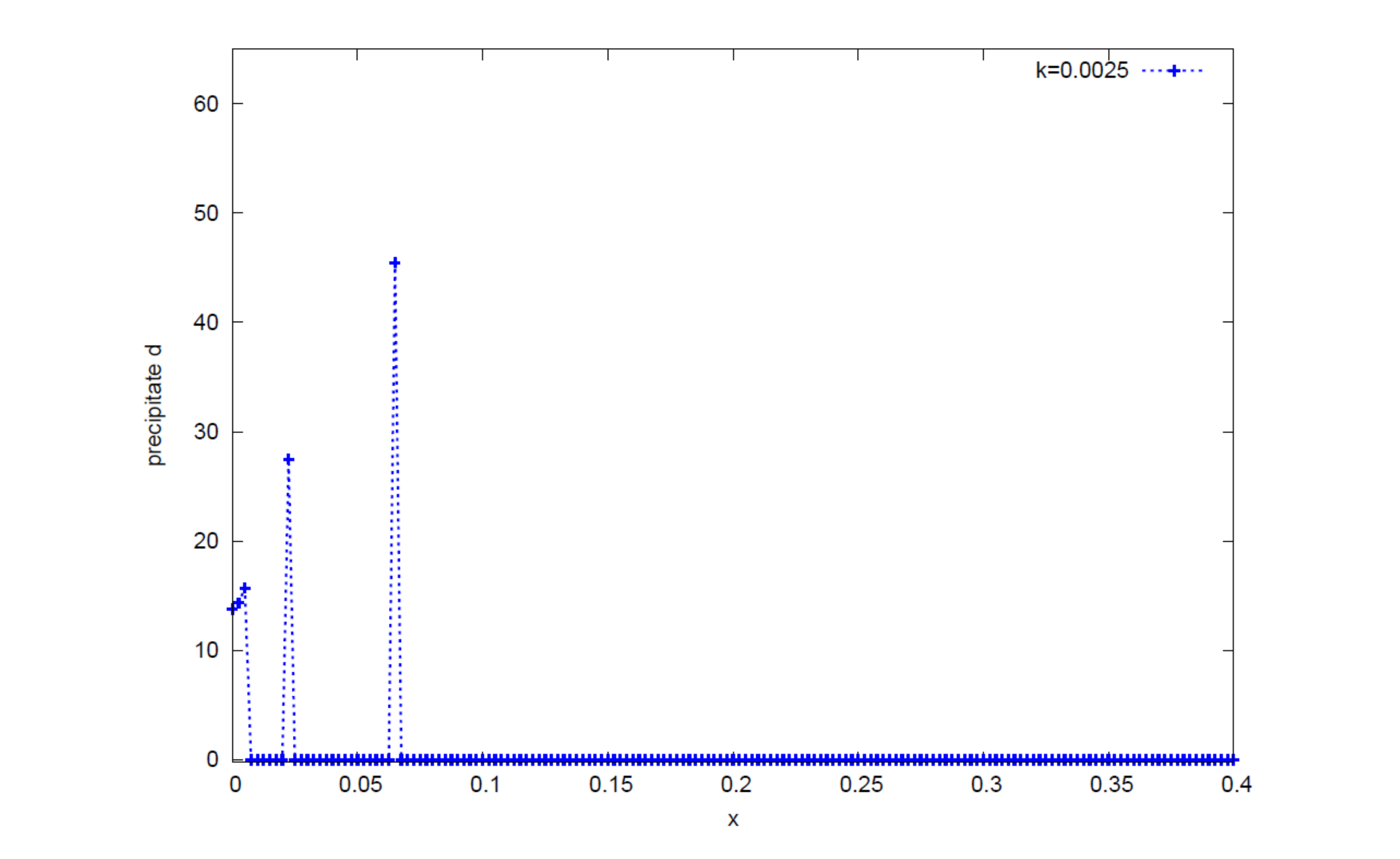}\quad
\includegraphics[width=3.6cm]{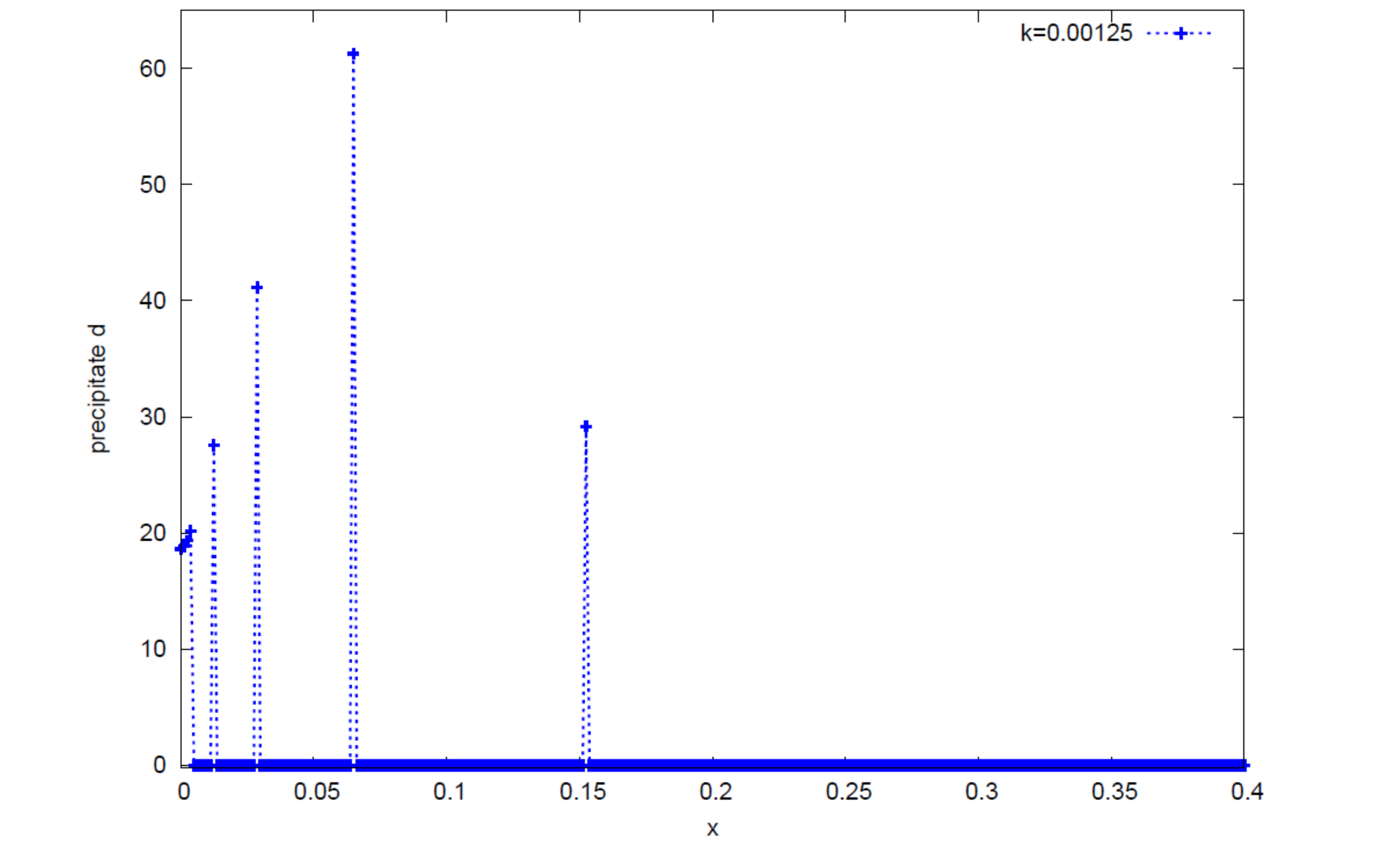}
\caption{Space concentration of precipitate substance $d$, for $\Delta x = 0.005, 0.0025$ and $0.00125$.} 
\label{fig:stab}
\end{figure}

Then we investigated numerically the stability of such model when the space step $\Delta x$ of the numerical grid changes. What we find is that the model is not stable, since as shown in Figure \ref{fig:stab} the bands change location, width and height when the grid spacing varies.

\vspace{0.5cm}

\paragraph{\bf Smoothing of the precipitation function.}

In order to overcome the problems occurring  in the numerical solution of the KR model, we will introduce
a smoothed version of the precipitation function $P(c,d)$ by removing some discontinuities in its definition. 
In fact, if we consider the two functions $ P(c,d)$ and $\frac{\partial}{\partial c} P(c,d)$,   they have discontinuity  
for $d=0$ and  $c = C^*$.   Moreover, the function $\frac{\partial}{\partial c} P(c,d)$ is discontinuous  for $c = C_s$ and for all $d > 0$.
The discontinuities can cause troubles in the computation of the numerical solutions of the system (\ref{KR}); 
for this reason the precipitation function (\ref{P}) has been redefined using the following  Heaviside smoothed function $H_{\sigma}(x)$,
having a smooth interval large $\sigma$: 
 \begin{equation}
\label{eq:heviside_smooth}
H_{\sigma}(x)= \left\{
  \begin{array}{ll}
    0             &  \mbox{if   } \: x  < 0 \\
    f(x)          &  \mbox{if   }  \: 0 \le  x \le \sigma \\
    1             &  \mbox{if   } \: x  > \sigma,   \\
  \end{array}
\right.
\end{equation}
where the function $f(x)$ possess all the following properties: it is defined in $[0, \sigma]$,  
$f(x)$ and $\frac{\mathrm d}{\mathrm d x} f(x)$ are continuous in $[0, \sigma]$, 
$f(0)=0$, $f(\sigma)=1$, 
$\frac{\mathrm d}{\mathrm d x} f(0) = \frac{\mathrm d}{\mathrm d x} f(\sigma)=0$
and $\frac{\mathrm d}{\mathrm d x} f(x) \ge 0$ for $x \in [0, \sigma]$.
The conditions adopted in the definition of $H_{\sigma}(x)$ ensure us that this function 
and its derivative are continuous in the all the points of the real line.

For example, a possible form of $f(x)$ is the following, for $x \in [0,\sigma]$:
 \begin{equation}
\label{eq:f_example}
f(x) = \left[ 1+ \sin \left( \pi \; \frac{x-\sigma/2}{\sigma} \right) \right] /2 
\end{equation}

Then the smoothed precipitation function can be written as follows:
\begin{equation}
\label{eq:precipitation_regular}
P_{\sigma}(c,d) = (c-C_s) \Big [ H_{\sigma}(c-C^*) +  H_{\sigma}(d) \; H_{\sigma}(c-C_s) \; (1-H_{\sigma}(c-C^*)) \Big ],
\end{equation}
which is defined by sums and products of continuous and derivable functions. 
Therefore, $P_{\sigma}(c,d)$
is continuous and, also, 
$\frac{\partial}{\partial c} P_{\sigma}(c,d)$,  
$\frac{\partial}{\partial d} P_{\sigma}(c,d)$ exist and are continuous too.

\begin{figure}[h]
  \centering
  \includegraphics[width=0.82\textwidth]{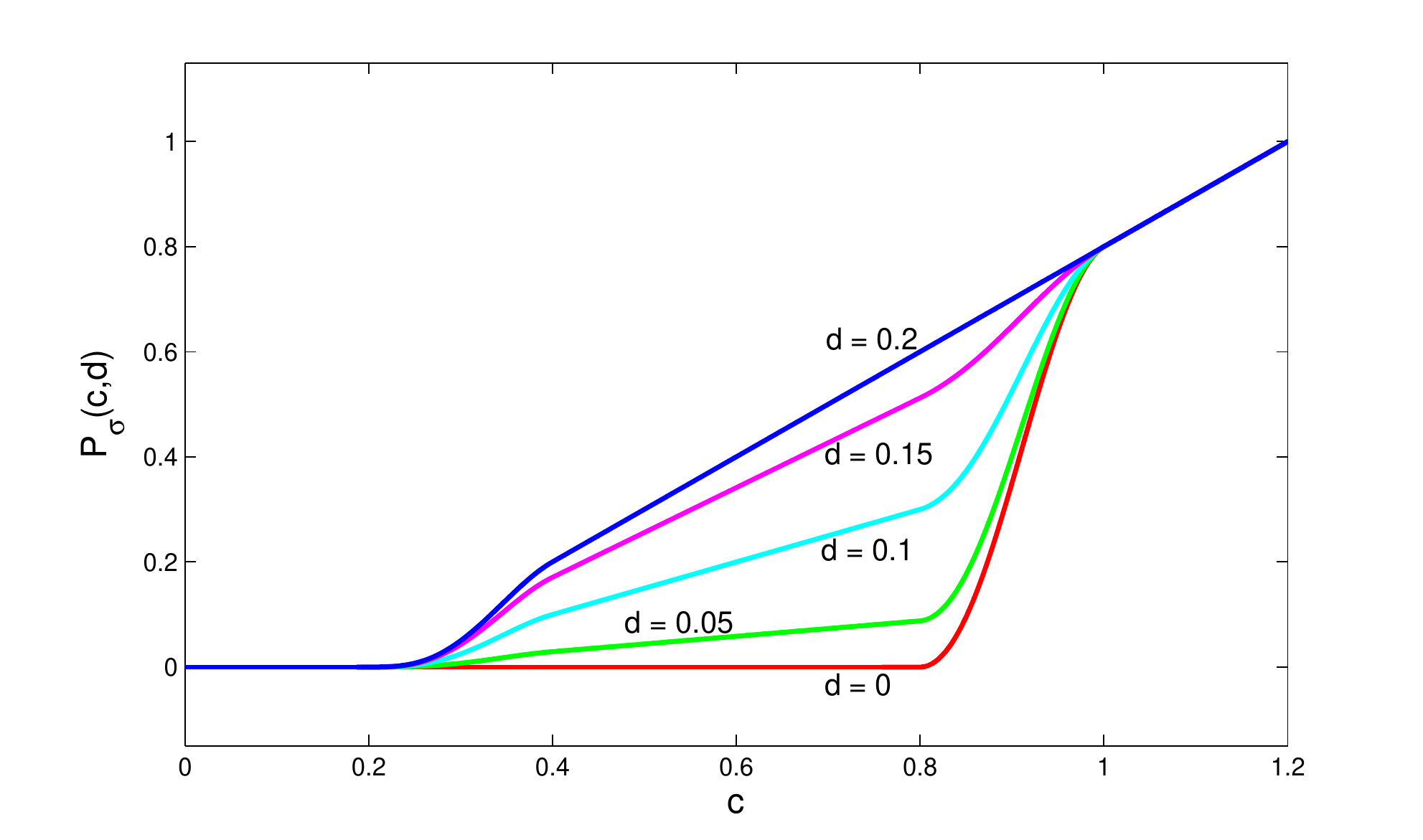}
\caption{Smoothed function $P_{\sigma}(c,d)$  
with  $\sigma= 0.2$, $C_s=0.2$, $C^*=0.8$. } \label{fig:smooth_P}  
\end{figure}

It is easy to to verify that, for $\sigma=0$, we have that $P(c,d)=P_{\sigma}(c,d)$.
For completeness, in Figure \ref{fig:smooth_P} the behavior of the smooth function $P_{\sigma}(c,d)$ for different values of $d$ is reported.

Using the finite element software COMSOL\textsuperscript{\textregistered}  5.2, some numerical experiments in a one-dimensional domain have been carried out using the smooth precipitation function $P_{\sigma}$, with the boundary and initial conditions (\ref{IB_conditions}) for system (\ref{KR}) and the parameters reported in Table \ref{table:param}. In particular, the COMSOL\textsuperscript{\textregistered} 
 5.2 internal {\it step function} with  smooth parameter $\sigma$ for $H_{\sigma}(x)$ in (\ref{eq:precipitation_regular}) has been used.
\begin{figure}[h]
\includegraphics[width=3.4cm]{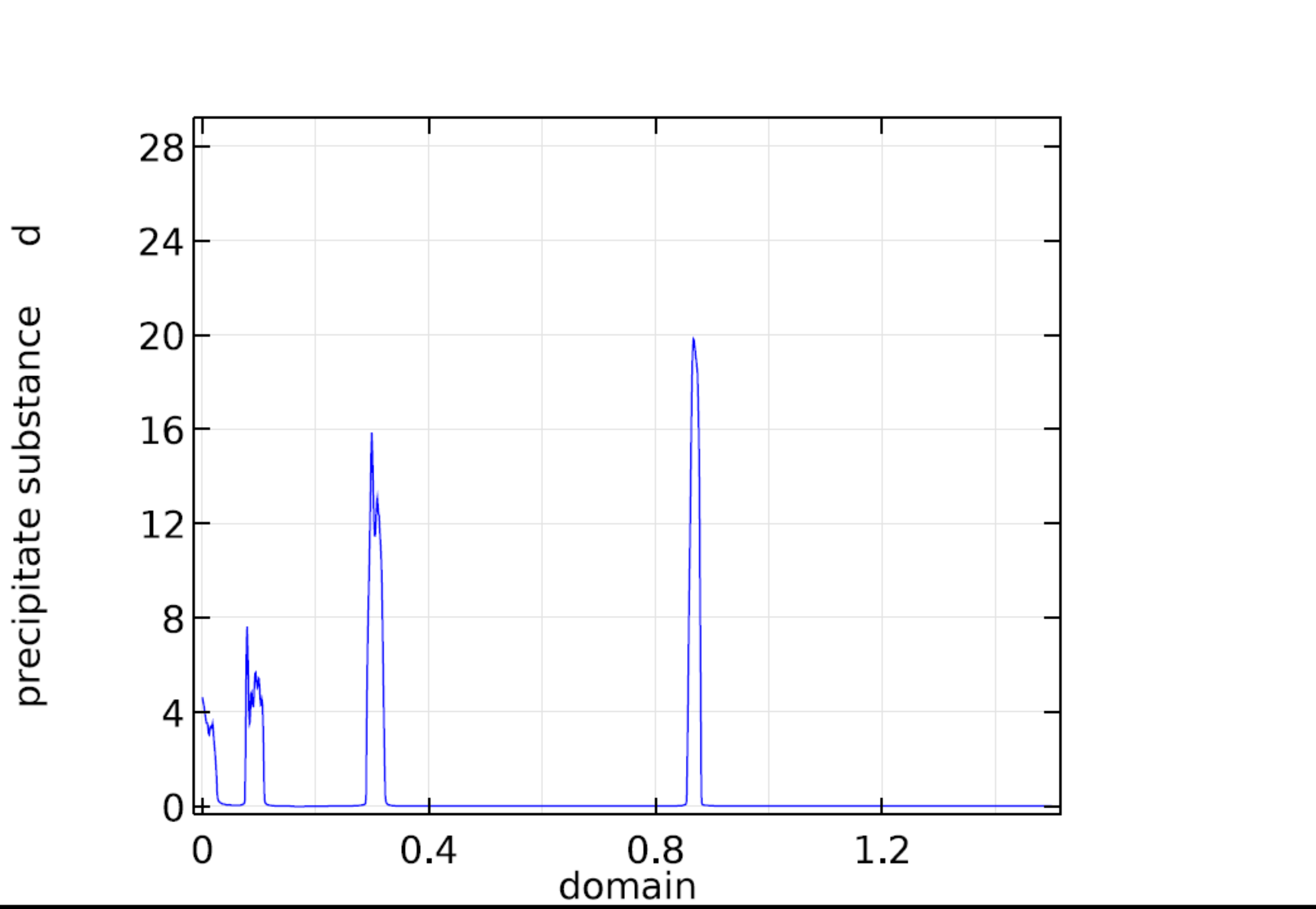}\quad
\includegraphics[width=3.4cm]{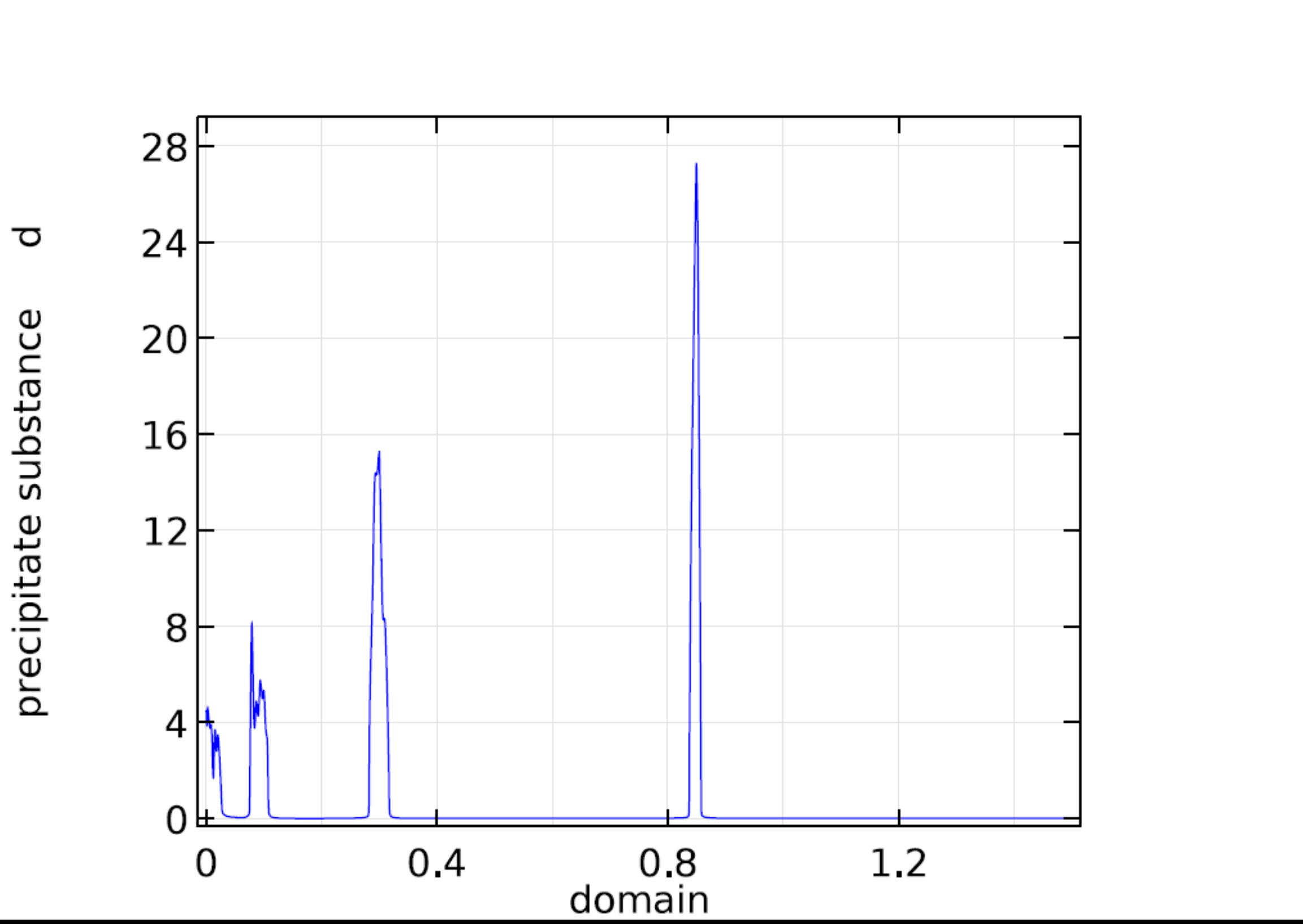}\quad
\includegraphics[width=3.4cm]{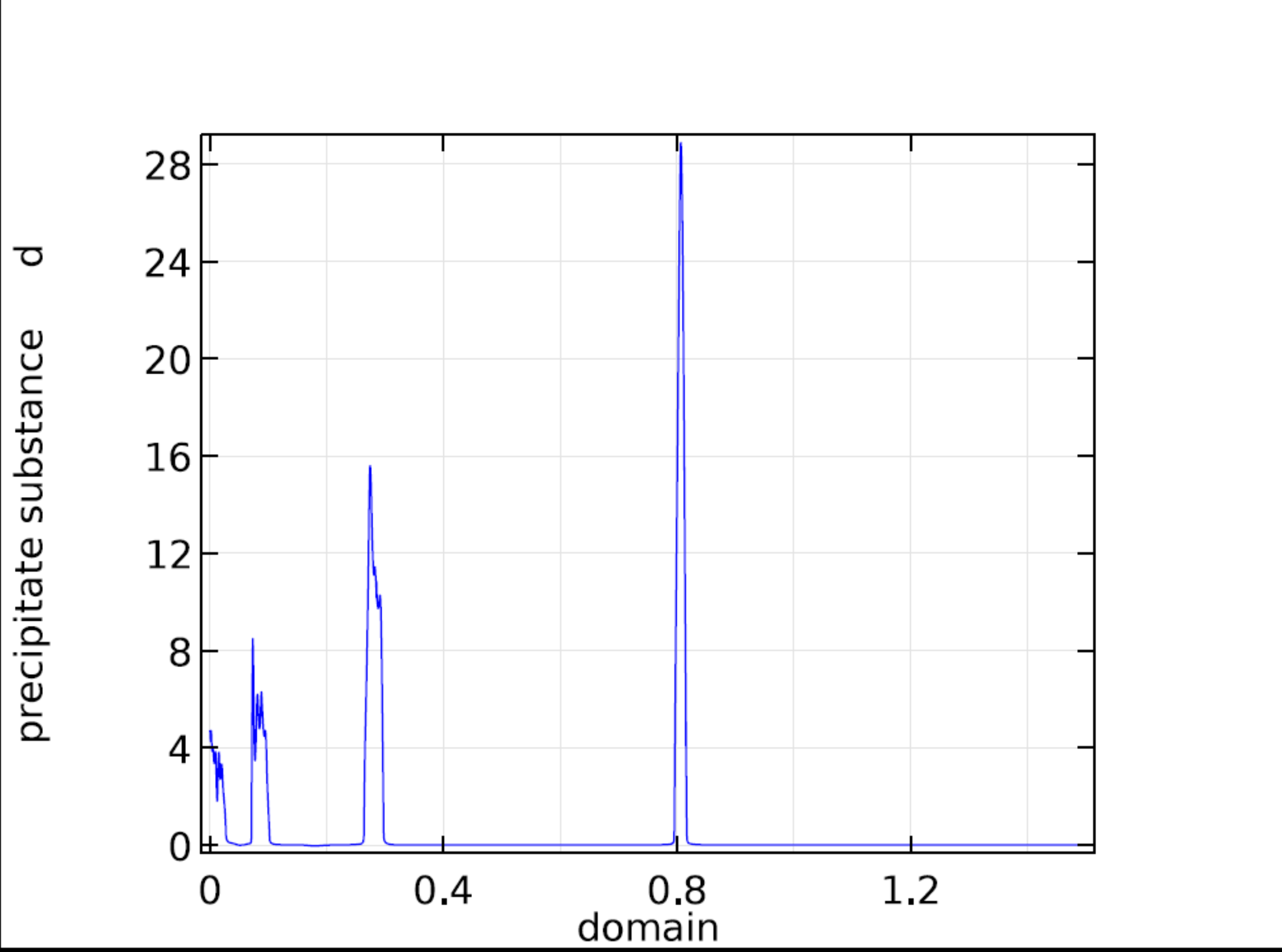}
\caption{Space concentration of precipitate substance $d$, for $\Delta x = 0.001, 0.0001$ and $0.00001$, and $\sigma=1.0$ in $P_\sigma$ defined in (\ref{eq:precipitation_regular}).} 
\label{fig:smoothKR_stability}
\end{figure} 

Numerical solutions of the problem (\ref{KR})-(\ref{IB_conditions}) in correspondence of  different values of the maximum size of the spatial mesh have been computed and the graphs in Figure \ref{fig:smoothKR_stability} show the spatial distribution of the precipitate substance $d$ for maximum size = 0.001, 0.0001 and 0.00001, respectively. Those results show that bands are numerically recovered and are quite stable in terms of location, intensity and width when the grid spacing varies, in contrast to what happens using the classical definition of the precipitation function (\ref{P}), see Figure \ref{fig:stab}. Hence, this outcome  suggests that the employment of a smooth precipitation function is necessary for computational and numerical purposes.
This conclusion makes it possible to advance the hypothesis that the use in the mathematical model of a smooth precipitation function could be more consistent with the physical reality of the phenomenon, instead of the classical discontinuous precipitation function.

\subsection{Test with KR model in two space dimension}\label{test_KR2D}

In this subsection, the results of the numerical simulation in a two dimensional  domain will be reported and compared with the outcomes of the chemical experiments.\\

\begin{figure}[h]
\begin{center}
\includegraphics[width=6cm,angle=-90]{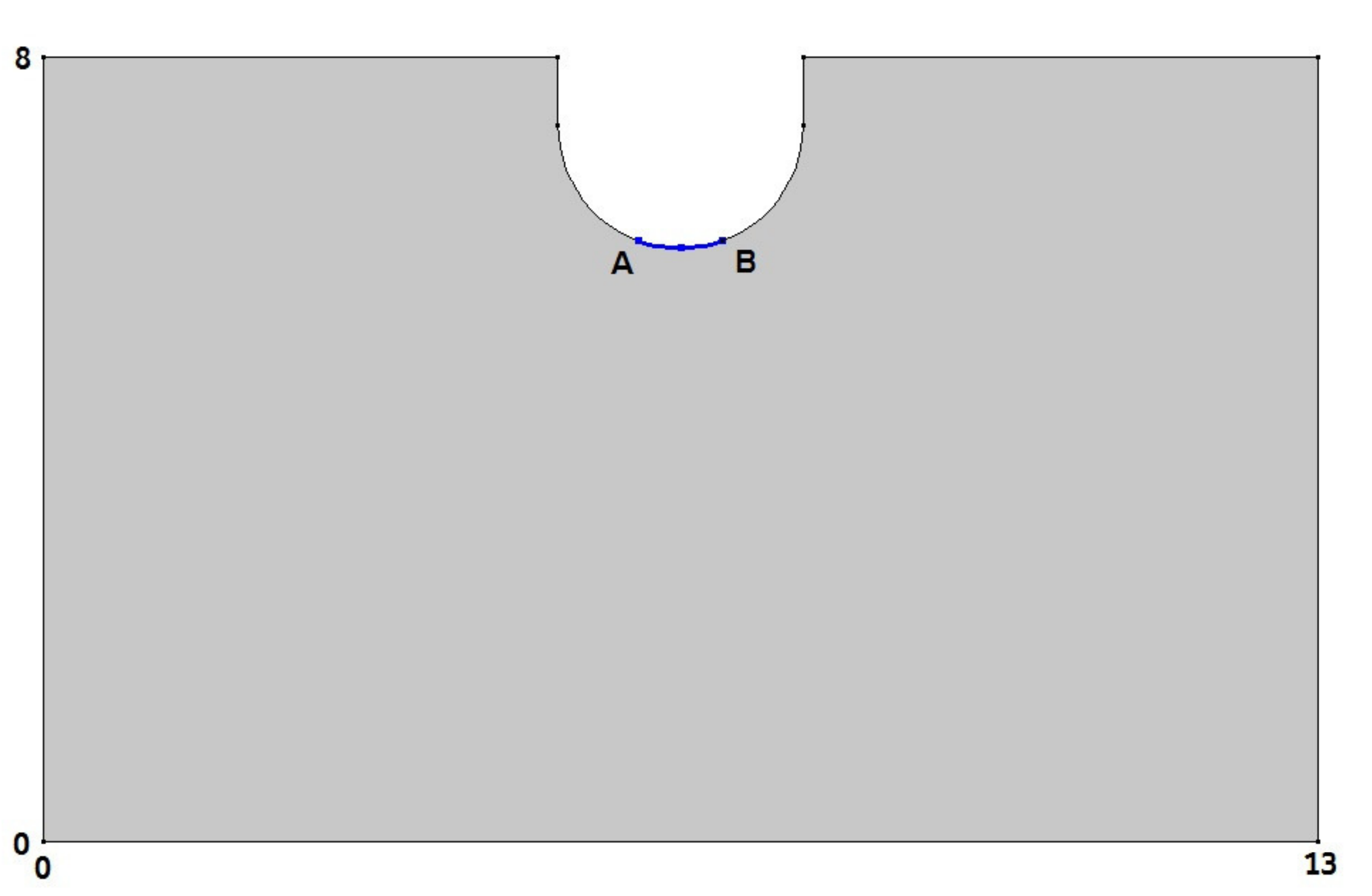}
\end{center}
\caption{Computational domain of experiment with Lecce stone.}
\label{fig:domain}
\end{figure}

\paragraph{\bf Results of numerical simulations: reproduction of Liesegang rings on Lecce stone}
In  Figure \ref{fig:rita} a zoom in, centered on the grooves, of two specimens of the laboratory experiment is shown and it correponds to an area of about 13$\times$8 $mm^2$.
According to the experimental setting described in Section \ref{sec:labexp}, the computational domain, reported in  Figure \ref{fig:domain} has the same dimensions: 13$\times$8 $mm^2$. The width of the groove is 2.5 mm and the center of the lower semicircle is horizontally centered and at a vertical distance of 0.7 mm from the upper boundary.

By examination of the pictures of the experimental Liesegang rings  obtained on the Lecce stone blocks, we can conclude that the presence, shape and position  of the rings are highly influenced by the physical contact between the stone and the iron nail. 
For this reason,  Dirichlet boundary conditions for the substance $a$  have been assigned only on a small portion of the groove, namely on the boundary arc $\wideparen{AB}$, as  shown in Figure \ref{fig:domain}. Denoting with $\Omega \subset \mathbb{R}^2$ the two-dimensional computational domain in Figure \ref{fig:domain} and with $\partial \Omega$ the boundary of  $\Omega$, then the initial and boundary conditions for  the unknowns $a(x,y,t)$, $b(x,y,t)$, $c(x,y,t)$, $d(x,y,t)$ of the system (\ref{KR}) are:

\begin{eqnarray}\label{conditions_rings}
 \begin{array}{lcll}
a(x,y,0)      & = &   0   & (x,y) \in \Omega, \\ 
b(x,y,0)      & = &   2   & (x,y) \in \Omega, \\
c(x,y,0)      & = &   0   & (x,y) \in \Omega, \\
d(x,y,0)      & = &   0  &  (x,y) \in \Omega, \\
a(x,y,t)       & = & 10   & (x,y) \in \wideparen{AB}, \; t>0, \\
a_x(x,y,t)   & = &    0       & (x,y) \in \{\partial \Omega \; - \wideparen{AB}\}, \; t>0,  \\
b_x(x,y,t)   & = &    0       & (x,y) \in \partial  \Omega, \; t>0\\
c_x(x,y,t)   & = &    0       & (x,y) \in \partial  \Omega,  \; t>0,\\
d_x(x,y,t)   & = &    0       & (x,y) \in \partial  \Omega,  \; t>0. 
 \end{array}
\end{eqnarray}

The differential problem (\ref{KR})-(\ref{conditions_rings})   
has been solved with the software COMSOL\textsuperscript{\textregistered} 5.2 
and the values of the parameters of the system (\ref{KR}) used in the simulations are reported in Table \ref{table:paramsint}.

\begin{table}[h]
\vspace{0.5cm}
\center
\begin{tabular}{l c l }
\hline
$D_a $  &  =  & 8.6806e-7   $ m^2  /  sec $\\ 
$D_b $  &  =  & 1.7361e-6   $m^2  /  sec $\\ 
$D_c $  &  =  & 1.7361e-6   $m^2  /  sec $\\ 
$k$  &  =  &  50   $m^3 sec^{-1} mol^{-1}$\\ 
$q$  &  =  &  100 $sec^{-1}$    \\ 
$ C_{s} $  &  =  & 0.2    $mol/m^3 $ \\ 
$ C^{*} $  &  =  & 0.8    $mol/m^3 $ \\ 
\hline
\end{tabular}
\vspace{0.5cm}
\caption{Parameters of the 2D simulations of the model (\ref{KR}) for the experiment with Lecce stone.}
\label{table:paramsint}
\end{table}
We stress that the parameters in Table \ref{table:paramsint} are not derived from a rigorous calibration with measured data, but they  have been chosen and adjusted, during the simulation,  in such a way to obtain  the rings. Actually, in the numerous numerical experiments carried out, it was observed that the problem is very sensitive to values of the parameters:
small variations of their values may cause a very different behavior of simulation results. For instance, if we scale of half order of magnitude the diffusion coefficients, the rings are lost. 

Therefore, numerical results suggest to investigate about the analytical characteristics of the differential problem such as: well-posedness, possible chaotic behavior, equilibrium points, spatial velocity of propagation  of the reactive front of the two substances $a$ and $b$ and so on.
If  the KR model well represents the physical phenomenon and if the high sensitivity of the problem to the coefficient values will be confirmed by the  qualitative analysis of the differential problem, then it will be possible to speculate that the formation of the rings, too,  is strongly dependent on the values of the parameters of the problem (\ref{KR})-(\ref{conditions_rings}). 
If this is the case, it will be possible to prevent the ring formation by changing some chemical or physical characteristic of the material, and consequently change the related parameters.\\
With the mentioned smoothing of the function $P$, we obtain the results shown in the next Figures: in the top graph of Figure \ref{fig:ab_t_4} the concentration in space of substances $a$ and $b$ at time  $t=4$ is depicted.
 Since substance $a$ enters the domain through the contact surface between the nail and the stone (namely the arc $\wideparen{AB}$), we translated this aspect in mathematical terms imposing Dirichlet boundary condition on $\wideparen{AB}$. We can observe that the substance $a$ is essentially present nearby  $\wideparen{AB}$ ; indeed it diffuses in the domain but, at the same time, it is also 
 consumed in the chemical reaction with $b$.

\begin{figure}[h]
\begin{center}
\includegraphics[width=7cm]{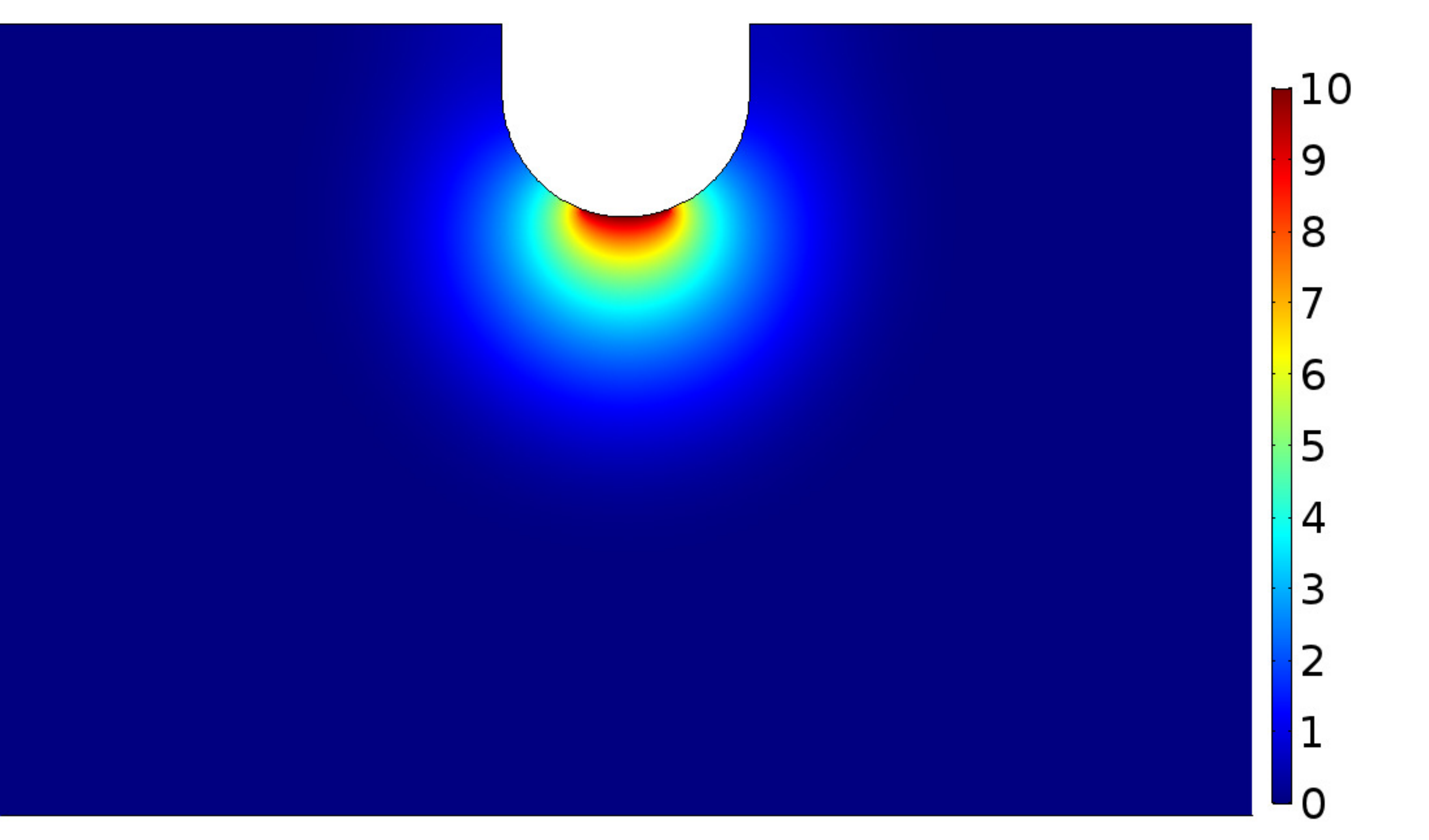}\quad 
\includegraphics[width=7cm]{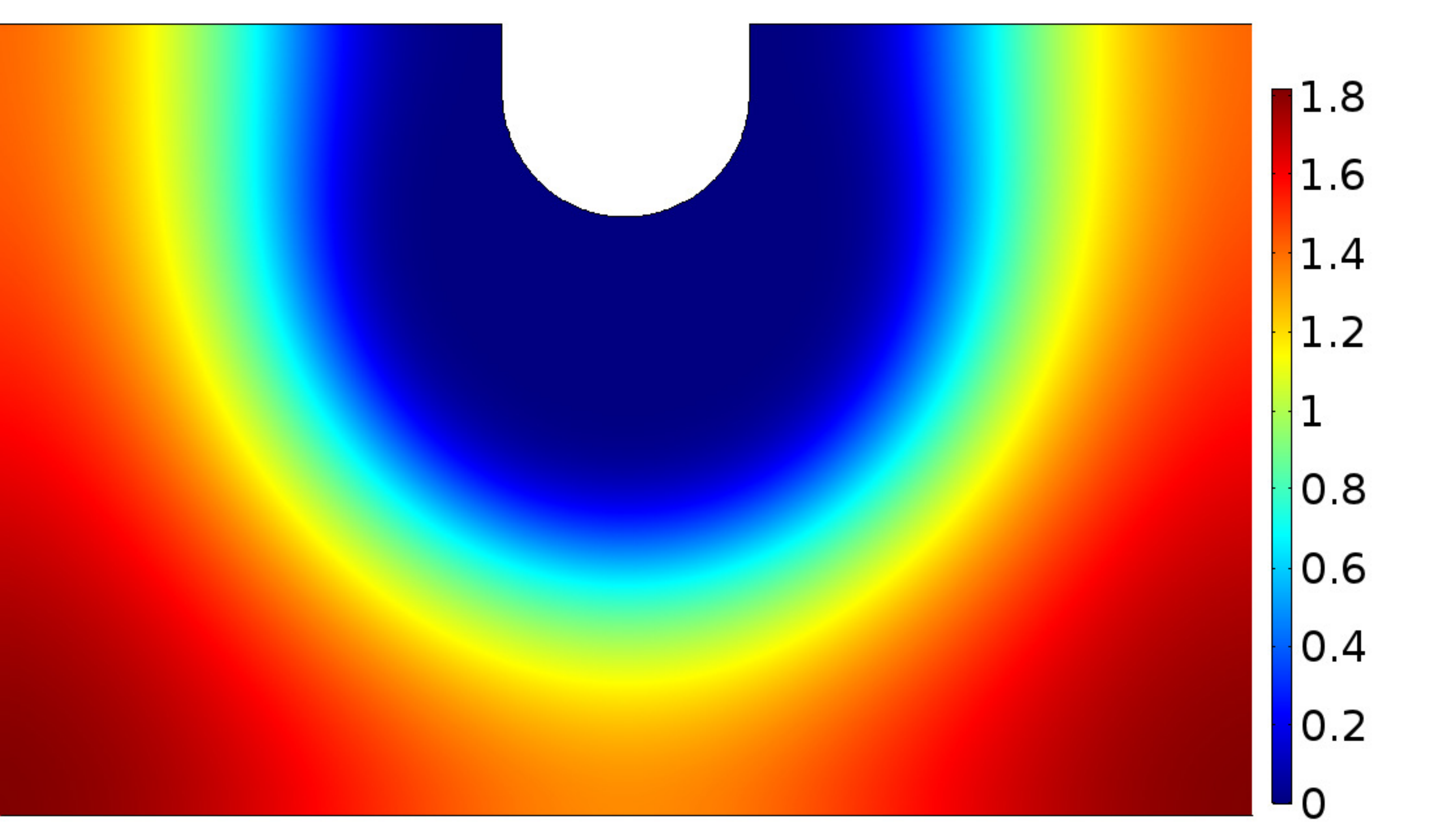}
\end{center}
\caption{Space concentration of substance $a$ (top) and $b$ (bottom) at time $t=4$.}  
\label{fig:ab_t_4}
\end{figure} 

\begin{figure}[h]
\begin{center}
\includegraphics[width=7cm]{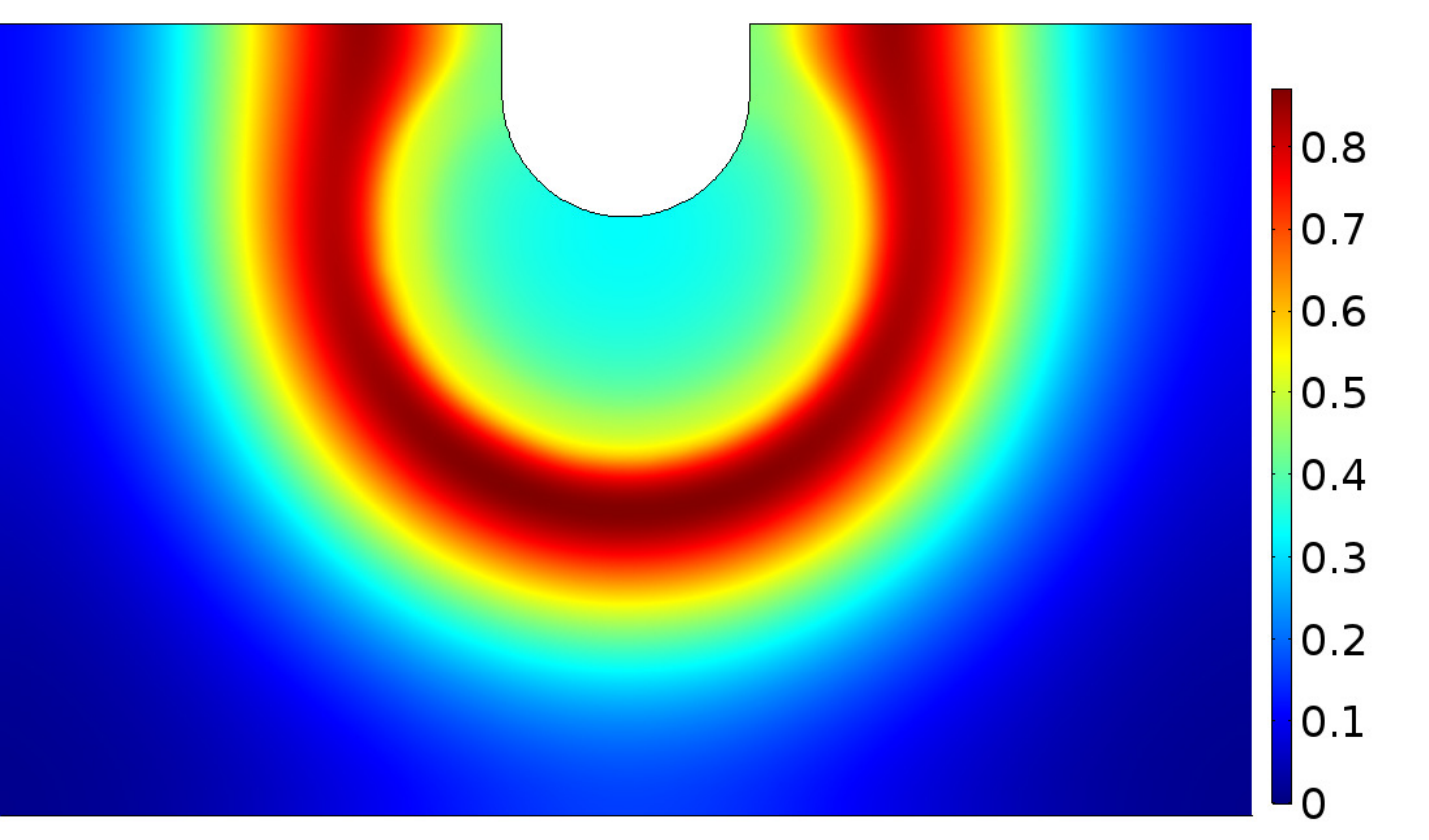}
\quad\includegraphics[width=7cm]{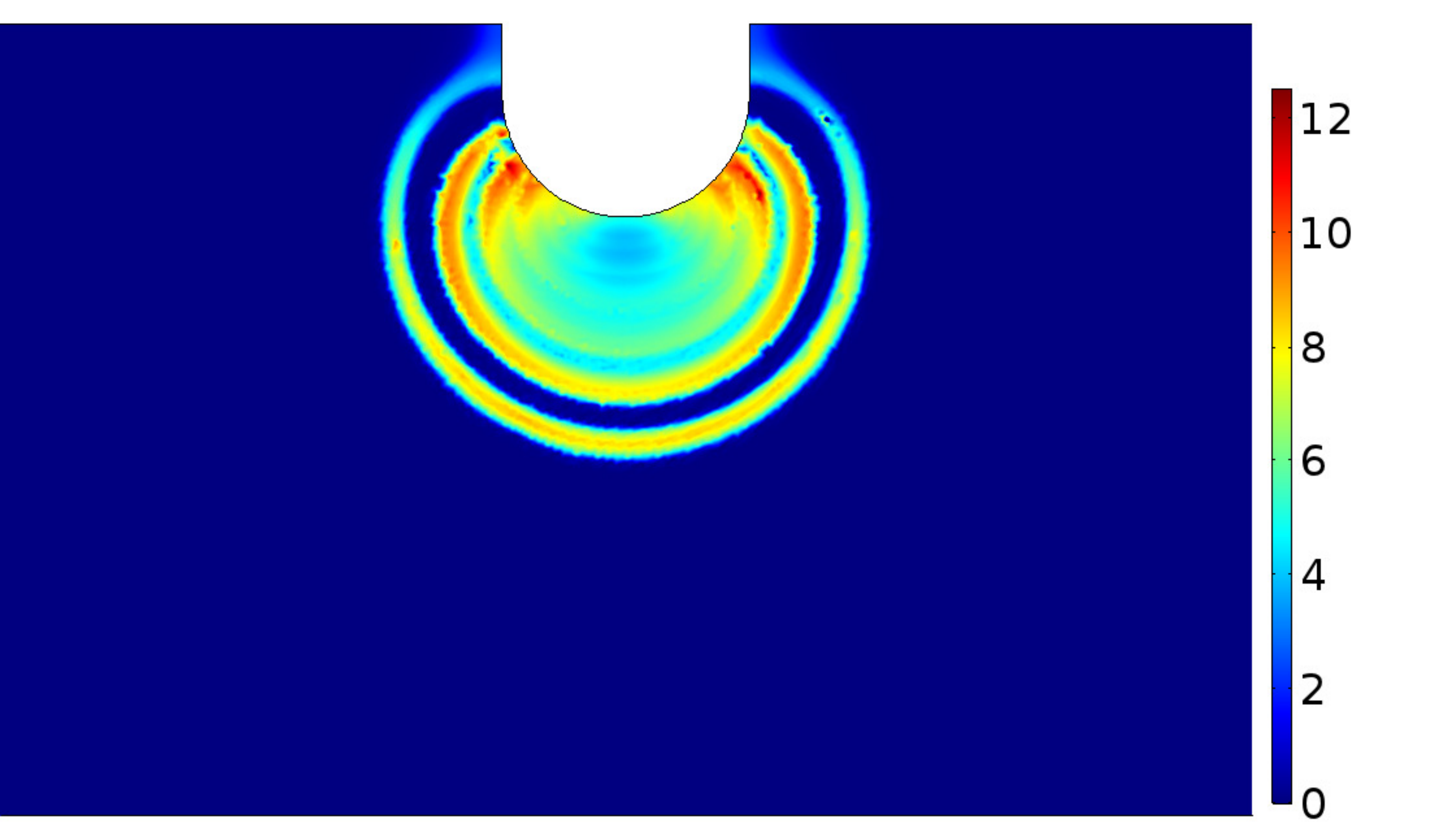}
\end{center}
\caption{Space concentration of substance $c$ (top) and $d$ (bottom) at time $t=4.0$.}  
\label{fig:cd_t}
\end{figure}

In the bottom graph of Figure \ref{fig:ab_t_4}, the concentration of substance $b$ at time  $t=4$ is depicted. At time $t=0$ this substance is uniformly distributed in the domain, that is: $b(x,y,0) = 2$, for $(x,y) \in \Omega$;  no ingoing or outgoing flux of $b$ is present on $\{\partial \Omega - \wideparen{AB}\}$ for $t>0$. Moreover, $b$  decreases while reacting with substance $a$ until it is completely consumed. 
Therefore, once the reaction occurs,  $b$ disappears and this  causes 
the end of the reaction itself. Note that the graph of $b$ shows that in the area where the reaction has occurred (see the graph of $d$ at time $t=4$ in Figure  \ref{fig:cd_t}) the concentration of $b$ is lower than the initial value or completely null.

The concentration of the substance $c$ at time $t=4.0$, is depicted on the top graph of Figure \ref{fig:cd_t}.
We can observe that the substance $c$, produced from the reaction between
$a$ and $b$, grows and advances in the spatial domain as time passes and
 subsequently it decreases, due to the precipitation process. 
The result is a peak wave enlarging in time, starting from the contact between the  nail  
and the stone. Moreover, behind the advancing peak, the value of $c$ is equal or below the saturation constant $C_s = 0.2$.
The bottom graph in Figure \ref{fig:cd_t} shows the concentration of the precipitate substance $d$ at time $t=4$ and we can observe the qualitative accordance with the chemical experimental rings. Indeed, during the numerical simulation, we can observe that the rings of the precipitate $d$ grow in number and the new generated rings are larger than the preexisting ones. 
This behavior is in accordance with phenomenon shown in Figure \ref{fig:provini}.
 Some animations of the evolution in time of the quantities under observation are on the webpage \cite{filmini}.\\
 
\begin{figure}[h]
\begin{center}
 \includegraphics[width=6cm,angle=0]{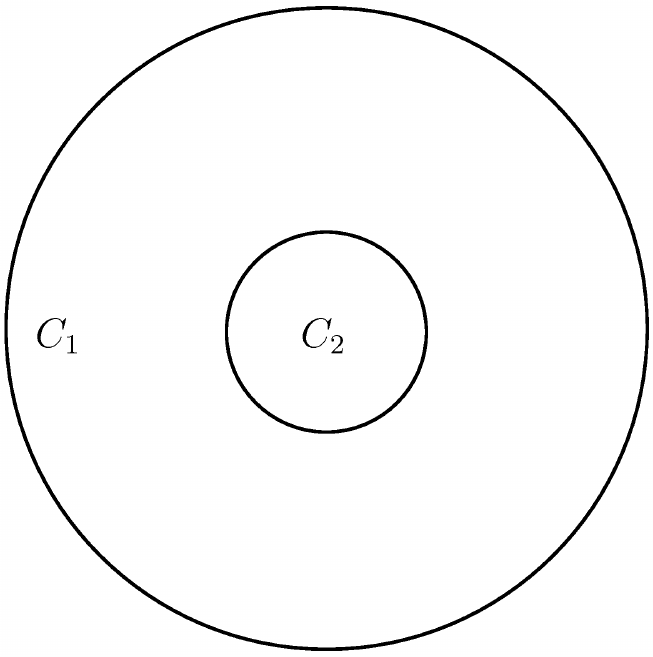}
\end{center}
\caption{Computational domain of experiment with agar gel.}
\label{fig:domain2}
 \end{figure}

\paragraph{\bf Results of numerical simulations: reproduction of Liesegang rings in agar gel.}
Here we set up the simulation parameters in order to reproduce the experiment described in subsection \ref{agar-agar}, see the images in Figure \ref{fig:agar}.
According to the experimental setting described in Section \ref{sec:labexp}, the computational domain, constituted by 2 concentric circles  as shown in  Figure \ref{fig:domain2} has dimensions of 60 $mm$ of diameter for the outer ring $C_1$ and 20 $mm$ of diameter for the inner circle $C_2$.

The initial and boundary conditions for  the unknowns $a(x,y,t)$, $b(x,y,t)$, $c(x,y,t)$, $d(x,y,t)$ of the system (\ref{KR}) are in this case:

\begin{eqnarray}\label{conditions_rings2}
 \begin{array}{lcll}
a(x,y,0)      & = &   0.0   & (x,y) \in C_1, \ a(x,y,0) =   0.1   \ (x,y) \in C_2, \\ 
b(x,y,0)      & = &   0.2   & (x,y) \in C_1, \ b(x,y,0)  =   0  \  (x,y) \in C_2, \\
c(x,y,0)      & = &   0   & (x,y) \in C_1 \cup C_2, \\
d(x,y,0)      & = &   0  &  (x,y) \in  C_1 \cup C_2, \\
a_x(x,y,t)   & = &    0       & (x,y) \in C_1 \cup C_2, \; t>0,  \\
b_x(x,y,t)   & = &    0       & (x,y) \in C_1 \cup C_2, \; t>0\\
c_x(x,y,t)   & = &    0       & (x,y) \in C_1 \cup C_2,  \; t>0,\\
d_x(x,y,t)   & = &    0       & (x,y) \in C_1 \cup C_2,  \; t>0. 
 \end{array}
\end{eqnarray}

The differential problem (\ref{KR})-(\ref{conditions_rings2}) has been solved with the software COMSOL\textsuperscript{\textregistered} 5.2 and, in the simulation, the values of the parameters of the system (\ref{KR}) are reported in Table \ref{table:paramsint2}.

\begin{figure}[h]
\begin{center}
\includegraphics[width=7cm]{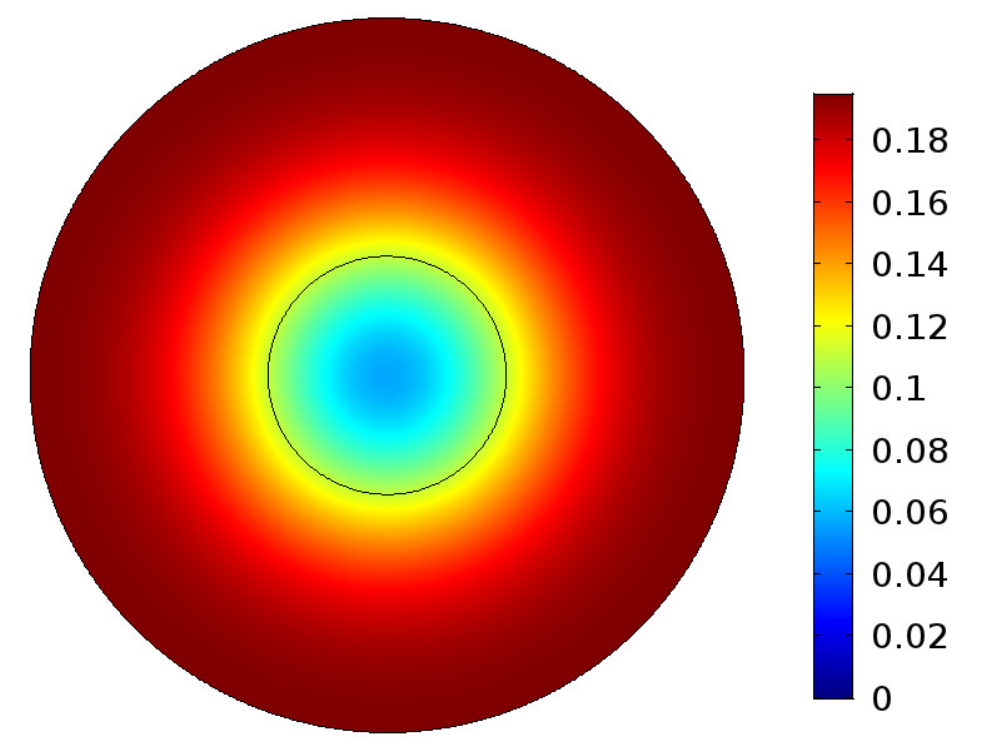}
\end{center}
\caption{Space concentration of substance $b$ at time $t=8$ hours.}  
\label{fig:bring_t_8}
\end{figure}

\begin{figure}[h]
\begin{center}
\includegraphics[width=7cm]{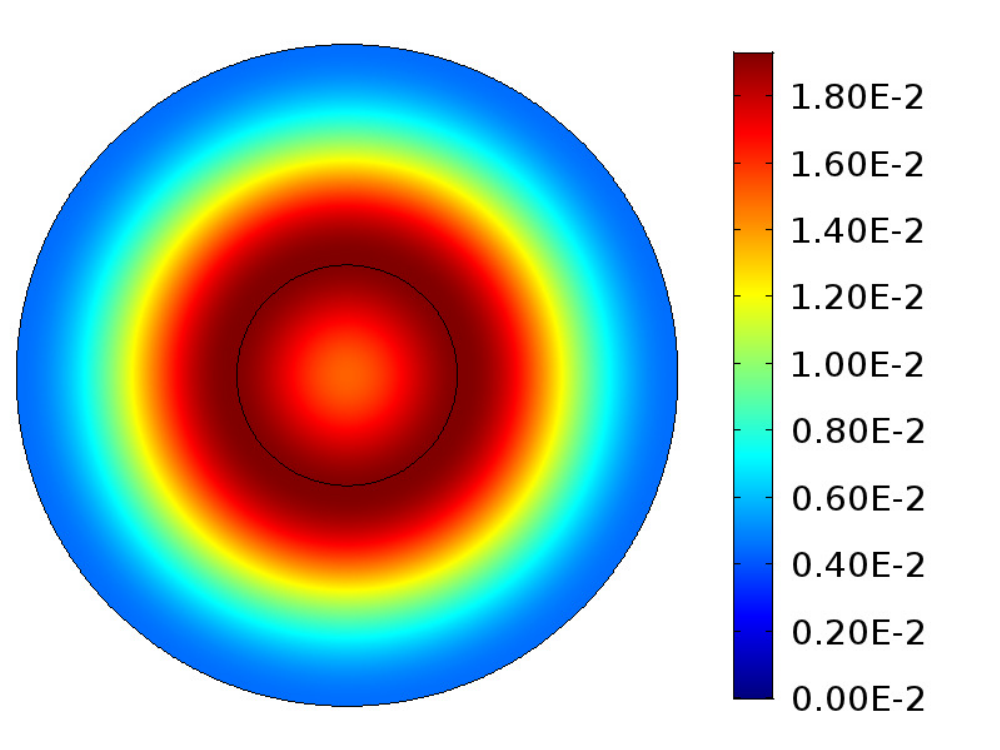}
\end{center}
\caption{Space concentration of substance $c$ at time $t=8$ hours.}  
\label{fig:cring_t_8}
\end{figure}

\begin{table}[h]
\vspace{0.5cm}
\center
\begin{tabular}{l c l }
\hline
$D_a $  &  =  & 1.7361e-7   $ m^2  /  sec $\\ 
$D_b $  &  =  & 8.6806e-10  $m^2  /  sec $\\ 
$D_c $  &  =  & 8.6806e-10  $m^2  /  sec $\\ 
$k$  &  =  &  0.0434   $m^3 sec^{-1} mol^{-1}$\\ 
$q$  &  =  &  85.81 $sec^{-1}$    \\ 
$ C_{s} $  &  =  & 0.02    $mol/m^3 $ \\ 
$ C^{*} $  &  =  & 0.04    $mol/m^3 $ \\ 
\hline
\end{tabular}
\vspace{0.5cm}
\caption{Parameters of the 2D simulations of the model (\ref{KR}) for the experiment with agar gel.}
\label{table:paramsint2}
\end{table}
It is worth noting that parameters in Table \ref{table:paramsint2} are not derived from laboratory experiments, but they  have been set during the simulation,  in such a way as to obtain  the Liesegang ring observed experimentally.\\
We report in Figure \ref{fig:bring_t_8} the concentration of substance $b$ at final time  $T=8$. We remark that the reaction between $a$ and $b$ and the consequent formation of $c$ lasts until substance $a$, initially present in the center of the domain (inner circle $C_2$), is not null and it stops as soon as $a$ vanishes. Substance $b$ is instead still present at final time T=8 (hours).
For the same reason, the vanishing of $a$ determines a stop in the precipitation process, since substance $c$ is neatly under the value $C_s=0.02 / mol/m^3$ and, consequently, also the growth of $d$ ends.
In the graph of Figure \ref{fig:cring_t_8}, the concentration of substance $c$ at final time  $T=8$ is depicted: the reaction product $c$ diffuses in the ring between $C_2$ and $C_1$ and  it finally decreases, due to the precipitation process.

\begin{figure}[h]
\begin{center}
\includegraphics[width=7cm]{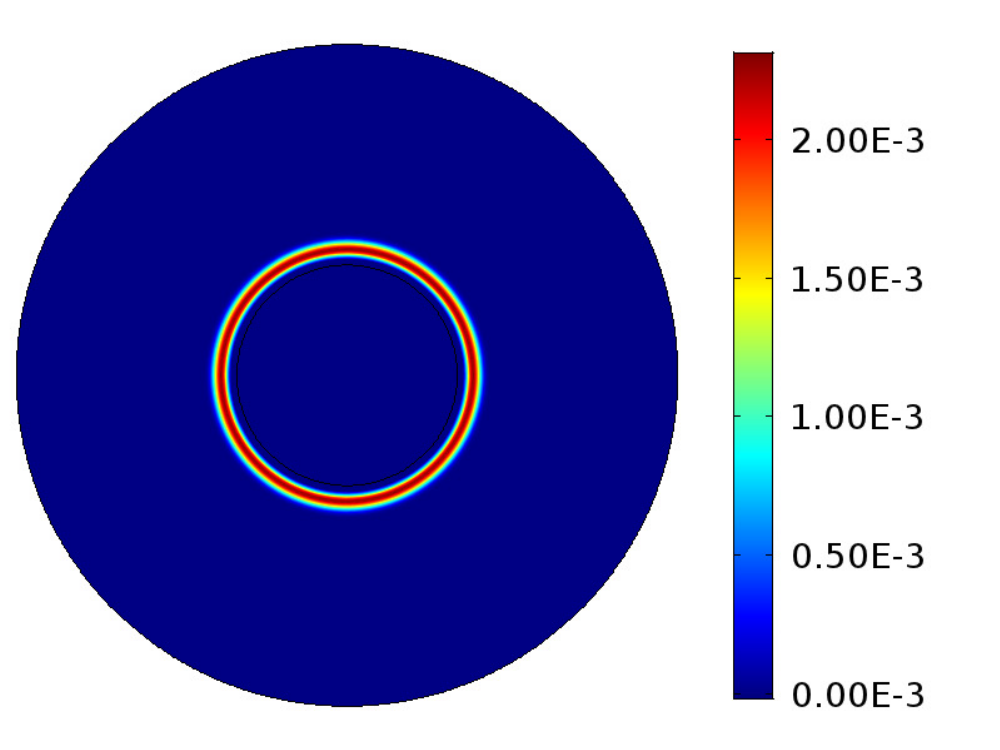}
\end{center}
\caption{Space concentration of substance $d$ at time $t=8$ hours.}  
\label{fig:dring_t_8}
\end{figure} 

In the graph of Figure \ref{fig:dring_t_8}, the concentration of precipitate substance $d$ at final time  $T=8$ is depicted, showing the formation of a ring of iron precipitate in the zone outside the inner circle $C_2$, as shown experimental outcomes, see Figure \ref{fig:agar}.
For the evolution in time of the quantities under observation, visit the webpage \cite{filmini}.

\section{Conclusions and perspectives}\label{sec:concl}

In this study on the formation of Liesegang rings we proposed a mathematical model to describe the process of precipitation of iron in Lecce stone. This model is a modification on Keller-Rubinow model where the precipitation function is replaced by a smooth function to avoid instabilities of the model. In such a way, we were able to develop a simulation algorithm capturing the main features of phenomenon of Liesegang formation in the stone: the shape and the intensity of rings depend on the nearness to the iron source, as shown in laboratory experiments. 
In the future, we aim at improving our mathematical model considering more aspects of this complex phenomenon, such as, for instance, the influence of water in the process of diffusion and crystallization of iron compound, but also the role of the pH concentration of the solution within the material.\\ Furthermore, we are interested in modeling the removal of iron stains not only from the surface but also from the inner layers of material.

\end{document}